\documentclass[10pt,journal,compsoc]{IEEEtran}

\ifCLASSOPTIONcompsoc
 
  \usepackage[nocompress]{cite}
\else
  \usepackage{cite}
 \fi
\usepackage{acronym}
\acrodef{KA}{Knowledge Area}
\acrodefplural{KAs}[KAs]{Knowledge Areas}
\acrodef{CyBOK}{Cyber Security Body of Knowledge}
\acrodef{CISSP}{Certified Information Systems Security Professional}
\acrodef{SSCP}{Systems Security Certified Practitioner}
\acrodef{CRISC}{Certified in Risk and Information Systems Control}
\acrodef{CISM}{Certified Information Security Manager}
\acrodef{CISA}{Certified Information Systems Auditor}
\acrodef{ISC2}[ISC\textsuperscript{2}]{Information System Security Certification Consortium, Inc}
\acrodef{ISACA} {Information Systems Audit and Control Association}
\acrodef{KwoPs}{Keywords or Phrases}
\acrodef{NCSC}{National Cyber Security Centre}
\usepackage{graphicx}
\graphicspath{{figures/}}
\usepackage{enumitem}
\usepackage{caption}
\usepackage{subcaption}
\ifCLASSINFOpdf
 
\else
 
\fi

\usepackage{amsmath}

\usepackage{url}
\usepackage{hyperref}
\usepackage{color,soul}
\usepackage{pdfpages}
\newcommand{\etal}[0]{et~al{.}}
\begin{document}

\title{A Framework for Mapping Organisational Workforce Knowledge Profile in Cyber Security} 

\author{Lata Nautiyal and Awais Rashid 
\IEEEcompsocitemizethanks{
\IEEEcompsocthanksitem Lata Nautiyal, University of Bristol, UK.\protect\\
E-mail: lata.nautiyal@bristol.ac.uk\\
Awais Rashid, University of Bristol, UK.\protect\\
E-mail: awais.rashid@bristol.ac.uk

\IEEEcompsocthanksitem University of Bristol, UK.\protect\\
}
}

\markboth{}%
{Shell \MakeLowercase{\textit{et al.}}: Bare Demo of IEEEtran.cls for Computer Society Journals}

\IEEEtitleabstractindextext{%

\begin{abstract}
A cyber security organisation needs to ensure that its workforce possesses the necessary knowledge to fulfil its cyber security business functions. Similarly, where an organisation chooses to delegate their cyber security tasks to a third-party provider, they must ensure that the chosen entity possesses robust knowledge capabilities to effectively carry out the assigned tasks. Building a
comprehensive cyber security knowledge profile is a distinct challenge; the field is ever evolving with a range of professional certifications, academic qualifications and on-the-job training. So far, there has been a lack of a well-defined methodology for systematically evaluating an organisation’s cyber security knowledge, specifically derived from its workforce, against a standardised reference point. Prior research on knowledge profiling across various disciplines has predominantly utilised established frameworks such as SWEBOK. However, within the domain of cyber security, the absence of a standardised reference point is notable. In this paper, we advance a framework leveraging \ac{CyBOK}, to construct an organisation’s knowledge profile. The framework enables a user to identify areas of coverage and where gaps may lie, so that an organisation can consider targeted recruitment or training or, where such expertise may be outsourced, drawing in knowledge capability from third parties. In the latter case, the framework can also be used as a basis for assessing the knowledge capability of such a third party. We present the knowledge profiling framework, discussing three case studies in organisational teams underpinning its initial development, followed by its refinement through workshops with cyber security practitioners.

\end{abstract}

}

\maketitle

\IEEEdisplaynontitleabstractindextext

\IEEEpeerreviewmaketitle

\IEEEraisesectionheading{
\section{Introduction}}

An organisation's effectiveness in mitigating risks is intrinsically linked to its capacity to manage pertinent knowledge \cite{neef2005managing}. The development of knowledge capability within an organisation is an ongoing process shaped by various activities. These activities include interactions among employees, various feedback processes, brainstorming sessions, sets of training, as well as acquiring new certifications or education \cite{ sandhawalia2011developing}. Knowledge capabilities are referred to as the abilities of an organisation to effectively utilise the acquired knowledge. These capabilities involve knowing when, where, and how to apply the knowledge within the organisation. This ability to initiate action from knowledge can stem from various sources and experiences \cite{ freeze2007knowledge}. Knowledge capabilities are also referred to as the ability of employees or organisations to recognise the value of new information, understand it, and use it to create new knowledge and skills, helping the organisation to grow \cite{gold2001knowledge }. 

Evaluating an organisation's knowledge capability involves conducting a comprehensive analysis of its workforce's knowledge capability \cite{lafayette2019learning}. This analysis takes into account their collective experiences, roles within the organisation, educational qualifications, certifications, training, and potentially other relevant factors. This rigorous assessment culminates in the provision of a comprehensive report, accompanied by tailored recommendations aimed at enhancing the organisation's readiness in the realm of cyber security.

Hence, in order to assess the knowledge capabilities, gaps, and strengths within the organisation, the concept of knowledge profiling has garnered significance. A knowledge profile is like a detailed map that shows the specific knowledge connected to a particular business process. It works as a guide, making it easier to find and use the most relevant information and expertise needed for job roles \cite{gourova2016knowledge}.

Security of information \& industrial systems in an ever-evolving threat landscape warrants a continuous supply of professionals equipped with state-of-the-art skills and knowledge \cite{kianpour2024more}. A nuanced understanding of the existing workforce is key to optimally augmenting aggregate organisational knowledge and expertise. However, in  the context of cyber security it is challenging given that cyber security roles and specialisations are often diverse -- no single academic degree or certification can claim to be suited to the needs of all roles\cite{dawson2018future}. The specialisms required for software security, for instance, vary significantly from those required of security operation centre workers. Furthermore, the workforce changes through employee departures, new arrivals or expansions. Even within a stable workforce, employees may gain new knowledge through on-the-job training, professional certifications or academic qualifications \cite{hart2020riskio}. Similarly, some knowledge may not remain current as an employee's current role may not require regular exercise and maintenance of such knowledge. Therefore, an organisation needs a systematic means to map its cyber security knowledge profile, as exhibited by the collective knowledge of its workforce, in order to identify areas of coverage,  strengths and where gaps may lie. Such a knowledge profile can enable an organisation to leverage its strengths through targeted investment in their maintenance and enhancement as well as to address the gaps via targeted recruitment, training or outsourcing, thus drawing in knowledge capability from third parties. In the latter case, organisations need a systematic way of assessing the knowledge profile of such third parties. To date, there are no frameworks available for systematically mapping organisational knowledge profiles against a common baseline. This absence not only limits organisations' ability to assess their capabilities but also hinders their capacity to determine their suitability for offering specific cyber security products and services \cite{dutton2019cybersecurity}.

Knowledge profiling in itself is not a new concept. It involves a systematic and effective method to ascertain where the knowledge sits and where the inherent gaps exist with respect to the environment\cite{wipawayangkool2019profiling}. In the context of cyber security, senior decision-makers, e.g., board members, may be interested in establishing if the organisational knowledge profile broadly meets the knowledge typically required for sectors in which it operates. Senior managers may be interested in coverage of particular types of knowledge, e.g., cryptography, software security, human factors or cyber-physical systems security. Team leaders, on the other hand, may wish to establish the depth of knowledge in such topics and specific topics required to ensure delivery of particular products or services. In other words, depending on the vantage point within an organisation, different views of a workforce knowledge profile may be required to deliver actionable intelligence at the correct level of abstraction.

In this paper, we propose such a framework which leverages \ac{CyBOK},\footnote{\url{https://www.cybok.org}}, as a baseline to map an organisation's knowledge profile. The framework has been developed through focused case studies in three organisations and has been refined based on two further workshops with cyber security practitioners. The framework develops a cumulative knowledge profile of a workforce by mapping to \ac{CyBOK} and compiling individual employees' cyber security knowledge based on their qualifications, certifications, on-the-job training and current roles. This knowledge profile can then be visualised at different levels of abstraction to suit the needs of different decision-makers within an organisation. It can also be used as a basis to demonstrate an organisation's knowledge capability when offering services as a third party. The framework also provides a means to demonstrate which knowledge is current and hence likely to be most up-to-date and which may be historical and may require a refresh.

The novel contributions of our work are threefold:

\begin{itemize}
\item We present a knowledge profiling framework to systematically map organisational knowledge profiles against a common baseline framework, \ac{CyBOK}. To our knowledge, ours is the first to propose such a framework.
\item We demonstrate how such a framework can help understand areas of focus and potential gaps through three case studies (which were used to develop the framework).
\item We highlight the framework's capability to provide insight at the correct level of abstraction for organisational decision-makers as well as its potential to demonstrate the portion of the knowledge profile being actively practiced within an organisation's functions. 
\end{itemize}

The rest of the paper is structured as follows. 
In Section~\ref{sec:background}, we present an overview of knowledge profiling, \ac{CyBOK} and its various use cases. 
Section~\ref{sec:profile} presents our knowledge profiling framework and how individual employees’ knowledge profiles are composed to derive an overarching knowledge profile. Section~\ref{sec:method} describes our methodological approach of developing and refining the framework, and dives deeper into our three case studies, highlighting how the framework provides an analytical lens on knowledge capability within organisations. It also discusses how the consultations with practitioners in our workshops led to further refinement of these analytical lenses. Section ~\ref{sec:related} provides an overview of the current state-of-the-art developments in the field. In Section ~\ref{sec:contri}, we highlight contribution of proposed work in terms of theory and practice.
Section ~\ref{sec:Disc} discusses the over-arching themes and insights emerging from our work while Section ~\ref{sec:conc} encapsulates our conclusions.

\begin{table*}
\centering
\fontsize{9}{9}\selectfont
\caption{\ac{CyBOK} Version 1.0.0 and Version 1.1.0 \acp{KAs} \emph{(Note that \ac{CyBOK} Introduction covers some basic definitions, concepts and principles but isn't deemed a standalone \ac{KA} in its own right even though it constitutes a Chapter in the full \ac{CyBOK})}}
\label{tab:ver}
\begin{tabular}{ |p{40mm} | p{56mm}| p{56mm}|} 
\hline

\vspace{1em}
\textbf{\ac{CyBOK} Broad category } \vspace{1em}&   
\vspace{1em}
\textbf{\ac{CyBOK} V1.0.0 Knowledge Areas} &
\vspace{1em}
\textbf{\ac{CyBOK} V1.1.0 Knowledge Areas} \\
\hline

Human, Organisational, \& Regulatory Aspects & Risk Management \& Governance  & Risk Management \& Governance\\[10pt]
\cline{2-3}
                & Law \& Regulation               & Law \& Regulation                                \\[10pt] \cline{2-3}
                               & Human Factors                   & Human Factors                                    \\ [10pt] \cline{2-3}
                               & Privacy \& Online Rights        & Privacy \& Online Rights                         \\ [10pt] \hline
Attacks \& Defences        & Malware \& Attack Technologies    & Malware \& Attack Technologies               \\ [10pt] \cline{2-3}
                               & Adversarial Behaviours          & Adversarial Behaviours                           \\ [10pt] \cline{2-3}
& Security Operations \& Incident Management & Security Operations \& Incident Management \\ [10pt] \cline{2-3}
                               & Forensics   & Forensics   \\ [10pt] \hline
Systems Security               & Cryptography                    & Cryptography                                     \\ [10pt] \cline{2-3}
                               & Operating Systems \& Virtualisation Security     & Operating Systems \& Virtualisation Security \\ [10pt] \cline{2-3}
                               & Distributed Systems Security    & Distributed Systems Security                     \\ [10pt] \cline{2-3}
 & Authentication, Authorisation \& Accountability & Authentication, Authorisation, \& Accountability   \\ [10pt] \cline{2-3}
                               & -                                &  Formal Methods for Security \\ [10pt] \hline
Software \&  Platform Security & Software Security               & Software Security \\ \cline{2-3}
                               & Web \& Mobile Security          & Web \& Mobile Security  \\[10pt] \cline{2-3}
                               & Secure Software Lifecycle       & Secure Software Lifecycle                        \\[10pt]  \hline
Infrastructure Security        & Network Security                & Network Security                              \\ [10pt] \cline{2-3}
                               &  Hardware Security              &  Hardware Security                               \\ [10pt] \cline{2-3}
                               & Cyber-Physical Systems Security &  Cyber-Physical Systems Security                             \\[10pt]  \cline{2-3}
                               & Physical Layer \& Telecommunications Security & Physical Layer \& Telecommunications  Security       \\ [10pt] \cline{2-3}
                                              & -              & Applied Cryptography           \\ [10pt] \hline

\end{tabular}
\end{table*}

\section{Background} \label{sec:background}

\subsection{Knowledge Profiling} 

The most important factor around which all strategies work for the growth of any organisation is \textit{'knowledge'} ~\cite{yao2023influence}. Organisations grow, remain competitive, and achieve their goals by leveraging their knowledge as a foundation ~\cite{kogut1992knowledge}. Every organisation is built using its fixed, financial and human resources. However, human resources play the most important role ~\cite{agarwala2003innovative}.
Hence, effectively managing human resources is an essential element for organisations to perform well and succeed. In any organisation, human resources drive value, promotion, and identity from their capabilities, which are shaped by their daily responsibilities, previous experiences, academic degrees, training, and certifications ~\cite{riddell2017role}. Evaluating employees' or human resources' expertise is essential for delegating tasks and identifying any gaps in the organisation~\cite{balderas2019model}.

Knowledge profiling is a mechanism for evaluating a variety of factors, including social networks, practices, tools, and capabilities pertinent to a team or a person or an organisation\cite{robin2019agent}. Therefore, assessing an organisation's knowledge, capabilities, gaps, and strengths has become crucial, highlighting the significance of knowledge profiling. \cite{ali2010conceptual}.

\subsection{\ac{CyBOK}} 

\ac{CyBOK} aims to capture the fundamental knowledge across key knowledge areas within cyber security. It aims to provide educators and trainers with a means to identify what is broadly accepted fundamental knowledge about a topic in the field with pointers to authoritative sources (as accepted by the community) for further details on particular concepts, approaches and methods. In that regard, \ac{CyBOK} is based on the premise that the knowledge exists out there in the form of research papers, text books, industry reports and standards. It, therefore, does not aim to \emph{create new knowledge but capture existing knowledge and only that which is deemed fundamental and broadly accepted as such by the community}. This resource has been developed over a period of approximately 4 years through extensive community engagement, both nationally in the UK and internationally. It involved more than $>$115 expert authors, reviewers, and advisors, along with over $>$1600 community comments. These efforts led to the incorporation of more than $>$2200 authoritative sources. 

The very first step was a nine-month scoping phase in 2017~\cite{rashid2018scoping} during which 19 \acp{KAs} (Table \ref{tab:ver}) were identified. This was followed by authorship and reviews of detailed \acp{KAs} texts by key experts in the field as well as public reviews. \ac{CyBOK} Version 1.0.0 (and associated resources) was released in October 2019. To ensure the currency of \ac{CyBOK} as a living document, \acp{KAs}  are always open to public review and new \acp{KAs} can also be proposed. Based on this indefinitely open call, a subsequent updated version \ac{CyBOK} V1.1.0 – containing two new \acp{KAs}  (Table \ref{tab:ver}) and a major revision to an existing \ac{KA} (Network Security) – was released in July 2021.

\ac{CyBOK} aims to establish a systematic, rigorous footing for foundational knowledge in the field of cyber security akin to more established relevant fields such as Software Engineering (e.g., SWEBOK~\cite{swebok}) and Computer Science (ACM computing curricula~\cite{acm_curricula}). It also complements existing frameworks, e.g., the ACM/IFIP/IEEE Joint Task Force on Cyber Security Curricula~\cite{jtf2017} and the NICE framework on Cyber Security Skills~\cite{nice}. \ac{CyBOK} captures the foundational knowledge that underpins such curricular guidelines or which is operationalised through skills frameworks such as NICE. \ac{CyBOK} has several key use cases, most critically providing an authoritative and common reference point for designing new university or professional training programmes. In addition, other use cases have emerged that utilise \ac{CyBOK} as a common reference point:

\textbf{New degree certification schemes.} One of the most extensive use cases in this regard has been the use of \ac{CyBOK} as a common reference framework to certify undergraduate and postgraduate degrees in cyber security at higher education institutions in the UK~\cite{nautiyal2022uk}. Programmes map the concepts covered within their respective modules to topics within \ac{CyBOK} and can demonstrate how they traceably meet the certification requirements for content coverage. This is done by utilising tree representations of the \acp{KAs} within \ac{CyBOK} which capture the relevant knowledge in a hierarchical format. Programmes need to show relevant coverage up to level 2 in suitable trees (note: there is no expectation that all \acp{KAs} within \ac{CyBOK} are covered; coverage depends on relevance as defined by the focus of a programme). 

\textbf{Contrasting the focus of different education or training programmes.} As any programme, whether university-level or professional training, can map its concepts to \ac{CyBOK}, this enables contrasting the focus of different programmes against a common baseline~\cite{hallett2018mirror}. It allows employers to understand if a particular programme will meet the knowledge needs of specific job roles (or how two programmes may contrast in that regard). On the other hand, it also enables learners to clearly distinguish if a programme will meet their knowledge acquisition goals. Such mappings follow a systematic process: 

\begin{description}
\item[Step 1] Key concepts covered in a programme are listed in the form of \ac{KwoPs}.
\item[Step 2] An extensive mapping reference available for each version of \ac{CyBOK} is used to identify in which \ac{KA}  such concepts may be found.
\item[Step 3] The mapping is established and logged by studying the relevant knowledge tree and, if needed, the detailed \ac{KA} text can be referenced.
\item[Step 4] If a suitable mapping isn't found through Step 2, then a summary of \ac{KA}  descriptions is used to identify which \ac{KA}  may be most suitable and the relevant knowledge tree and \ac{KA} text is studied to establish the mapping.
\end{description}

\begin{figure*}[ht]
  \centering
  \begin{subfigure}[b]{0.3\textwidth}
    \includegraphics[width=\textwidth]{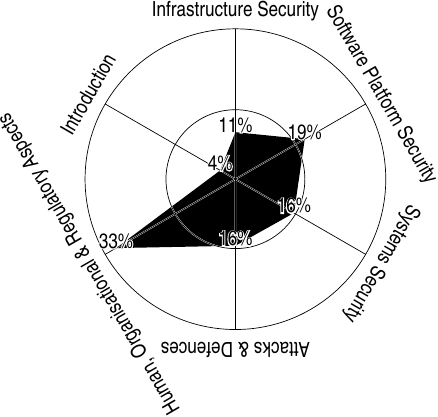}
    \caption{Mapping of \ac{CISSP} to \ac{CyBOK} Broad Categories}
    \label{fig:cissp_broad}
    \end{subfigure}
    \hspace{2cm}
    \begin{subfigure}[b]{0.35\textwidth}
     \centering
    \includegraphics[width=\textwidth]{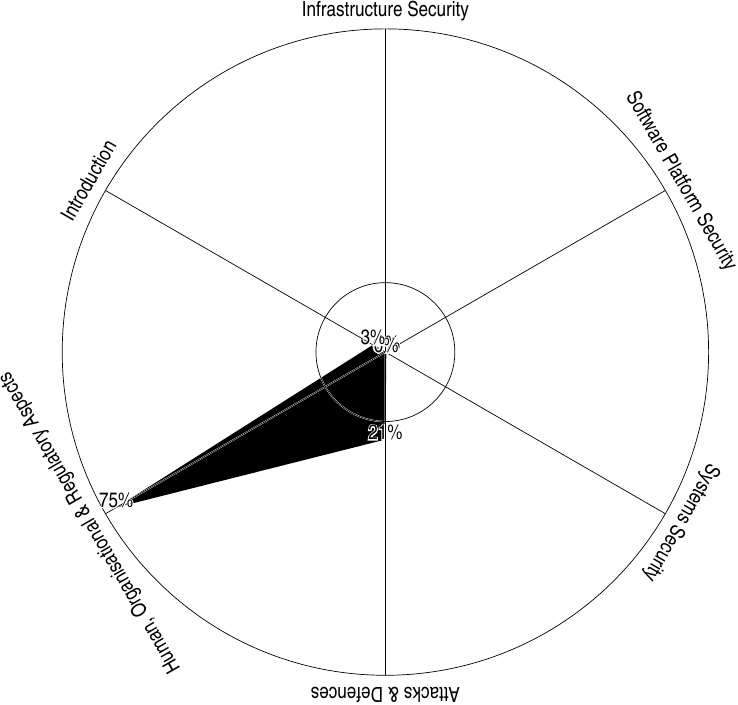}
    \caption{Mapping of \ac{CISM} to \ac{CyBOK} Broad Categories}
    \label{fig:cism_broad}
    \end{subfigure}
    \caption{Contrasting \ac{CISSP} and \ac{CISM} in terms of coverage of \ac{CyBOK} Broad Categories}
\end{figure*}

\begin{figure*}[ht]
  \centering
    \begin{subfigure}[b]{0.4\textwidth}
        \centering
        \includegraphics[width=\textwidth]{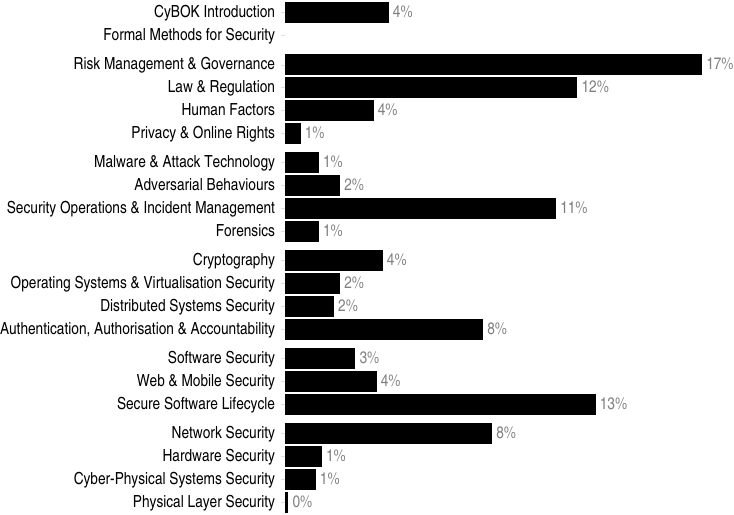}
        \caption{Mapping of \ac{CISSP} to \ac{CyBOK} \acp{KAs}}
        \label{fig:cissp_perKA}
    \end{subfigure}
    \hspace{2cm}
    \begin{subfigure}[b]{0.4\textwidth}
        \centering
        \includegraphics[width=\textwidth]{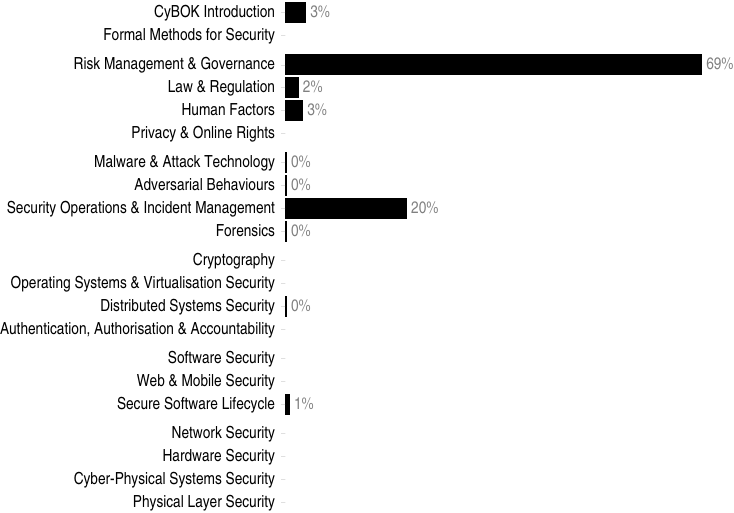}
        \caption{Mapping of \ac{CISM} to \ac{CyBOK} \acp{KAs}}
         \label{fig:cism_perKA}
    \end{subfigure}
    \caption{Contrasting \ac{CISSP} and \ac{CISM} in terms of coverage of \ac{CyBOK} \acp{KAs}}
\end{figure*}

These mappings enable one to contrast the focus of different programmes at different levels of granularity, e.g.,  Figures~\ref{fig:cissp_broad} and ~\ref{fig:cism_broad} show how two professional certifications  a \ac{CISSP} from (ISC)\textsuperscript{2} and \ac{CISM} from \ac{ISACA} -  contrast in terms of their coverage of the \ac{CyBOK} broad categories in Table~\ref{tab:ver} while Figures~\ref{fig:cissp_perKA} and  Figure~\ref{fig:cism_perKA} enable one to observe the degree of coverage of different \ac{CyBOK} \acp{KAs}. 

\vspace{1em}
\fbox{\begin{minipage}{24em}
These comparisons clearly demonstrate the variation in coverage between the broad categories of \ac{CyBOK}.
In Figure~\ref{fig:cissp_broad}, Human, Organisational, \& Regulatory Aspects cover 33\%, whereas in Figure ~\ref{fig:cism_broad}, they cover 75\%. This shows that \ac{CISSP} focus is spread across almost all broad categories, although Human organisational and regulatory aspects dominate. On the other hand, \ac{CISM} clearly indicates a much stronger focus towards Human organisational and regulatory aspects. 
\end{minipage}}
\vspace{1em}

Note that these mappings utilise \ac{CyBOK} Version 1.0.0 -- the current version at the time the mappings were conducted. We utilise these within our knowledge profiling framework discussed next and hence utilise \ac{CyBOK} Version 1.0.0 as a basis of that framework. This is in line with best practiced in such bodies of knowledge where historical versions are maintained to ensure that it is clear which version is being utilised for a particular use case and that users do not have to immediately transform their existing programmes as a new version materialises (and such updates can happen in line with any update processes for programmes utilising such bodies of knowledge).

A notable study ~\cite{hallett2018mirror} systematically evaluated four major frameworks: IISP ~\cite{IISP2017}, JTF~\cite{BURLEY2017}, NICE (NIST SP 800–181) ~\cite{NEWHOUSE2017}, and \ac{NCSC} Certified Master’s ~\cite{NCSC2017} with \ac{CyBOK}. \ac{CyBOK} acted as the baseline for the comparison. \textit{(IISP can be seen as the predecessor of \ac{CyBOK} for \ac{NCSC} certified degree programmes)}. This evaluation looked at the coverage of these frameworks among the \ac{CyBOK} broad categories of Attacks \& Defences, Human, Organisational and Regulatory Aspects, Infrastructure Security, Software \& Platform Security and Systems Security and also the \ac{CyBOK} \acp{KAs}. The study reported that IISP focused more on attacks \& defences but has a low coverage of systems security, software \& platforms security and infrastructure security. NICE had similar results to IISP with respect to attacks \& defences but with comparatively more focus on the areas with low coverage by IISP. JTF and \ac{NCSC} Certified Master’s covered a wider range of cyber security topics across all the broad categories. All the four frameworks demonstrated considerable coverage of human, organisational \& regulatory aspects.

This study highlights that \ac{CyBOK} can serve as a foundation for comparing the focus of various academic programmes, acting as a baseline. It illustrates the relative coverage of individual programmes and can be utilised for assessing the coverage of different professional certifications. Moreover, these professional certifications provide a valuable foundation for constructing the knowledge profile of an organisation. \ac{CyBOK} has been a baseline for the \ac{NCSC} certified programmes on a national level. These certified programmes are documented and listed in a \ac{CyBOK} mappings booklet, providing a comprehensive overview of the programmes \cite{CyBOKBooklet}.

\begin{figure}
\centering
\includegraphics[width=0.49\textwidth]{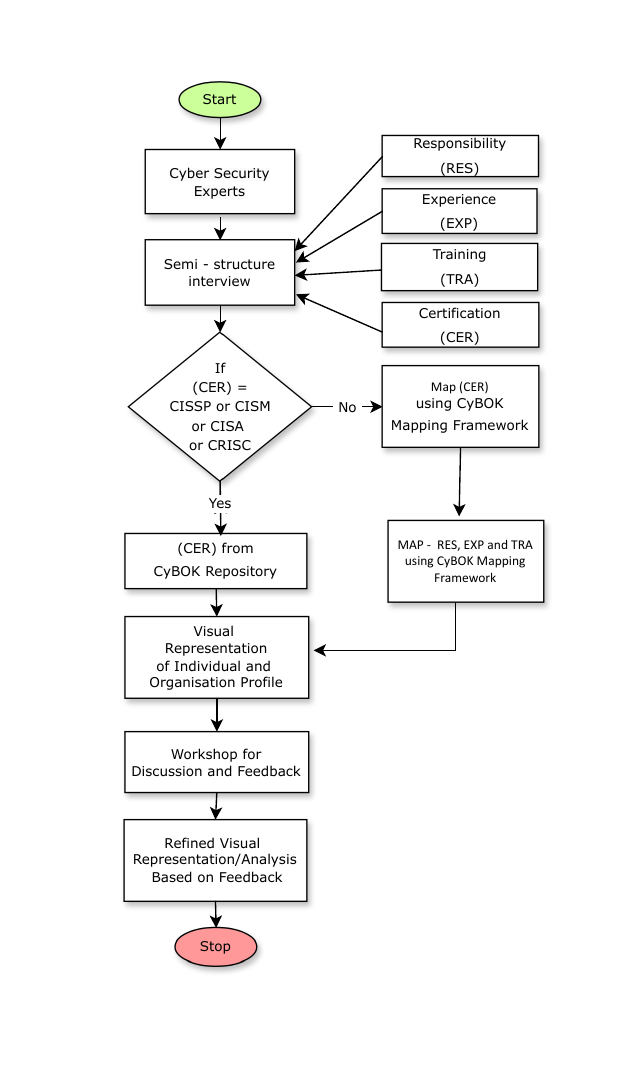}
\caption{Methodology Flow Chart}
\label{fig:BOX_LINE}
\end{figure}

\vspace{1em}
\fbox{\begin{minipage}{24em}
\ac{CyBOK} consists of 21 \acp{KAs}, with new areas added as they emerge through the change management process. Each \ac{KA} provides in-depth knowledge of a specific domain. The \ac{CyBOK} Introduction covers introductory topics, such as definitions and principles, which are typically included in introductory units in degree programmes. Although it cannot be a separate \ac{KA} therefore not a part of  (Table \ref{tab:ver}), it is included in mappings because of its importance in covering these essential definitions and principles.

\end{minipage}}
\vspace{1em}

\section{Knowledge Profiling Framework}
\label{sec:profile}
The knowledge profiling framework is rooted in development of an employee's individual knowledge profile and then taking a union of the knowledge profiles of the whole team/organisation to derive a cumulative profile. An employee's individual knowledge profile is characterised by:

\begin{description}

\item[CER] The academic or professional certifications they hold, e.g., academic qualifications such as a Bachelor's or Master's in Cyber Security or a professional qualification such as \ac{CISSP} \cite{cissp}, \ac{CISM} \cite{cism}.
\item[TRA] Training they may have undertaken, e.g., on-the-job or through additional courses.
\item[EXP] Knowledge acquired through experience in cyber security roles in their career.
\item[RES] Knowledge acquired through responsibilities in the current role.

\end{description}

\textbf{Formal Specification -}
\begin{enumerate}
\item \textbf{\ac{CyBOK} (CYB) Definition:} 
\ac{CyBOK} (CYB) is the collection of all the knowledge areas (KA). It can be defined as:\\
$CYB=\{KA_1,KA_2,KA_3,....KA_{n}\}$ \\
Where 'n' represents the number of knowledge areas within \ac{CyBOK}. 

\item \textbf{Knowledge of an Employee Definition:} The knowledge possessed by an employee is a set 'K', which consists of four distinct elements i.e. CER, TRA, EXP, and RES. The set 'K' is formally defined as: \\
\[K =  \{CER, TRA, EXP, RES\}\] 

\item \textbf{Concept Definition:} The knowledge components (CER, TRA, EXP, RES) are structured in the form of concepts, which are represented as sets of phrases or keywords. These concepts are formally defined as:
\\
$C= \{C_1, C_2, C_3...C_n\}$\\

Where 'n' represents the number of concepts identified within each of the four knowledge components (CER, TRA, EXP, RES). The process of identifying these concepts is performed individually for each knowledge component.

\item \textbf{Mapping of Concepts to \ac{CyBOK}:} All the identified concepts are mapped onto \ac{CyBOK} (CYB) using a generic function called Map. This Map function performs the mapping operation by taking a concept (C) as input and returns the results in the form of triplets denoted as $<KA, t, d>$. Each triplet comprises the following elements:

KA: The knowledge area to which the concept C is mapped.

t: The specific topic within the knowledge area.

d: The depth or level of depth within the mapped knowledge area.

Mathematically, this mapping operation can be represented as:
\\
$C_{<KA,t,d>}\leftarrow Map(C, CYB)$

Where:
\\ $C_{<KA,t,d>}$ \\ denotes the result of mapping concept C onto \ac{CyBOK}.

\item	\textbf{Mapping of Knowledge Components (K) over \ac{CyBOK} (CYB):} The mapping of knowledge components (K) onto the \ac{CyBOK}(CYB) can be formally defined as follows:
\\${K \over {CYB}} \leftarrow\forall_{1 \le x \le c} map (C_x, CYB)$ \\

Where:
\\${K \over {CYB}}$ \\ represents the set of knowledge components mapped onto \ac{CyBOK}.
x denotes an index ranging from 1 to the total number of concepts (c) identified. Cx represents the x-th concept. 

\item	\textbf{Knowledge Profile of the Employee:} The knowledge profile of an employee $KP_{EMP}$\\ is established as the union of the knowledge components (K) possessed by the employee. 

Mathematically, it is defined as:

$KP_{EMP} \leftarrow \left[ \bigcup\ {K \over {CYB}} \right]$\\

\item	\textbf{Knowledge Profile of the Organisation:} The Knowledge Profile of the Organisation is denoted as $KP_{ORG}$  and is formally characterised as the union of all individual employee knowledge profiles, represented as $KP_{EMP}$, that collectively constitute the organisational knowledge profile. It can be defined as:
\[KP_{ORG} = \bigcup \{KP_{EMP1}, KP_{EMP2}, \ldots KP_{EMPn}\}\]

The union operation is used to combine these individual employee knowledge profiles into the knowledge profile of the entire organisation. This signifies that the organisation's knowledge profile encompasses the knowledge possessed by all individual employees, forming a collective representation of the organisational knowledge profile.

\end{enumerate}

\textit{Note that the process of deriving these unions is automated as all mappings are maintained in a machine readable format (CSV) and a union can be undertaken as well as the visualisations derived automatically. The mappings are a manual process. However, as more certifications and qualifications are mapped to \ac{CyBOK}, a ready bank of such mappings makes this a less resource intensive process. During our case studies (see Section~\ref{sec:method}) we found that, in comparison to TRA and CER, the number of concepts that required to be mapped from EXP and RES was rather modest.}

\begin{itemize}
\item For organisation A the total number of concepts covered in the two categories for TRA and CER was 94.10 percent, but only 5.07 percent for EXP and RES. 

\item For organisation B the total number of concepts covered in the two categories for TRA and CER was 73.52 percent and 26.47 percent for EXP and RES.

\item For organisation C the total number of concepts covered in the two categories for TRA and CER was 87.52 percent and 12.48 percent for EXP and RES.
\end{itemize}

Next, we will elaborate on our approach for developing and enhancing our knowledge profiling framework, and we will also provide a comprehensive overview of the insights we acquired from our case studies.

\begin{figure*}[ht]
  \centering
  \begin{subfigure}[b]{0.4\textwidth}
    \includegraphics[width=\textwidth]{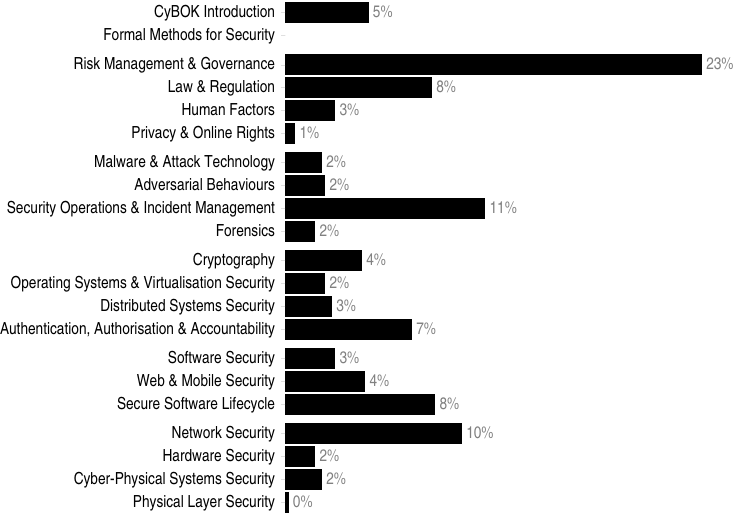 }
    \caption{ Histogram of Employee X }
    \label{fig:HistoX}
    \end{subfigure}
    \hspace{3cm}
\begin{subfigure}[b]{0.4\textwidth}
    \includegraphics[width=.7\textwidth]{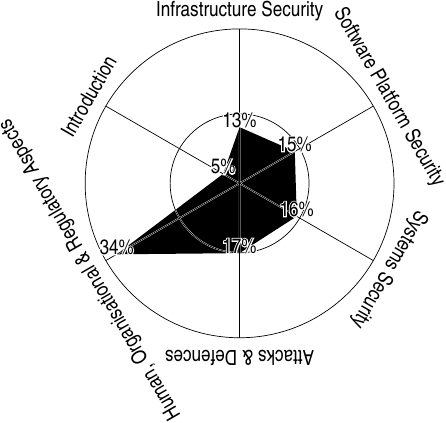}
    \caption{Spider Diagram of Employee X }
    \label{fig:spiderx}
    \end{subfigure}
    \hspace{3cm}
    \caption{Histogram and Spider Diagram of Employee X to show the \ac{CyBOK} Knowledge Area and \ac{CyBOK} Broad Category Coverage}
\end{figure*}

\begin{figure*}[ht]
  \centering
  \begin{subfigure}[b]{0.4\textwidth}
    \includegraphics[width=\textwidth]{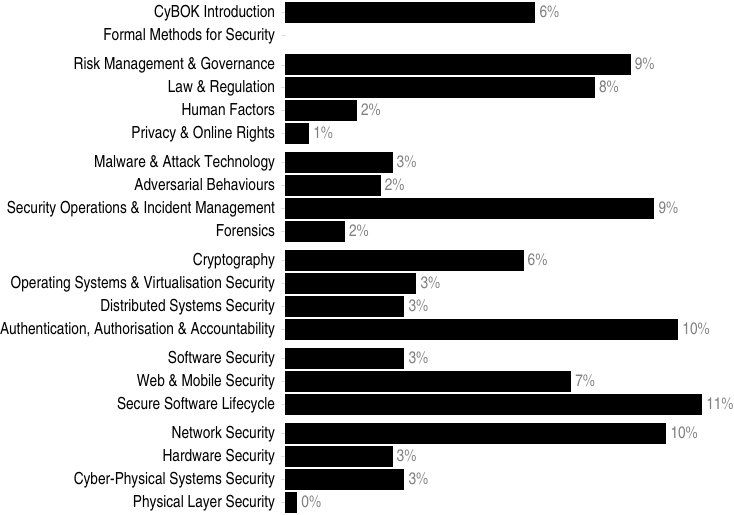 }
    \caption{ Histogram of Employee Y }
    \label{fig:HistoY}
    \end{subfigure}
    \hspace{3cm}
\begin{subfigure}[b]{0.4\textwidth}
        \includegraphics[width=.7\textwidth]{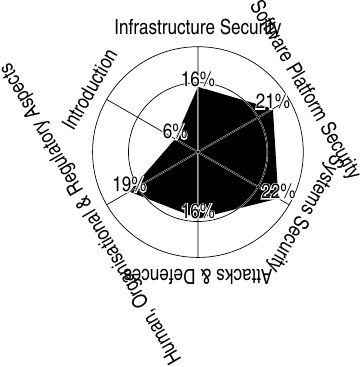}
    \caption{Spider Diagram of Employee Y }
    \label{fig:spidery}
    \end{subfigure}
    \hspace{3cm}
    \caption{Histogram and Spider Diagram of Employee Y to show the \ac{CyBOK} Knowledge Area and \ac{CyBOK} Broad Category Coverage}
\end{figure*}

\begin{figure*}[ht]
  \centering
  \begin{subfigure}[b]{0.4\textwidth}
    \includegraphics[width=\textwidth]{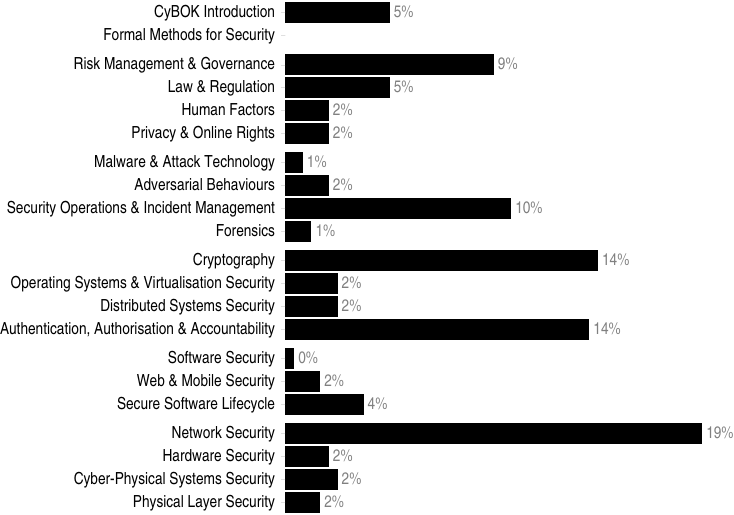 }
    \caption{Histogram of Employee Z}
    \label{fig:HistoZ}
    \end{subfigure}
    \hspace{3cm}
\begin{subfigure}[b]{0.4\textwidth}
            \includegraphics[width=.7\textwidth]{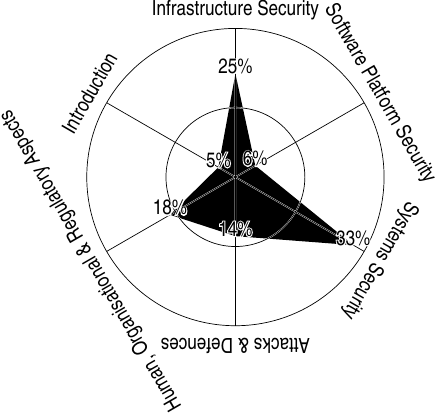}
    \caption{Spider Diagram of Employee Z}
    \label{fig:spiderz}
    \end{subfigure}
    \hspace{3cm}
    \caption{Histogram and Spider Diagram of Employee Z to show the \ac{CyBOK} \ac{KA}  and \ac{CyBOK} Broad Category Coverage}
\end{figure*}

\section{Methodology}
\label{sec:method}

To convert unstructured data into organised elements, we adhere to the procedure outlined by ~\cite{creswell2017research}, ~\cite{seaman1999qualitative}, which provides detailed instructions for conducting empirical studies. This process helps us systematically categorise and analyse the data, ensuring clarity and consistency in our findings. One significant advantage of using qualitative methods is that they compel the researcher to delve deep into the complexity of the issue rather than simplifying it. This approach ensures that the results are more detailed and informative, providing a richer understanding of the subject matter. We undertook the research in two distinct phases, \textbf{Case study} and \textbf{Workshop}. The case study phase served as a foundational step in developing the initial knowledge profiling framework.  The workshops were conducted to assess the framework's efficacy, engaging stakeholders for feedback and refining the framework iteratively based on their valuable insights and suggestions. This collaborative process ensured that the final framework was robust, practical, and aligned with the needs and perspectives of key stakeholders.

\vspace{1em}
\fbox{\begin{minipage}{24em}
\textbf{Note :} We provide a concise overview of our methodology in Figure \ref{fig:BOX_LINE}. For a thorough, step-by-step understanding, refer to the detailed explanation in the appendix, illustrated in Figure \ref{fig:FlowChart}. 

\end{minipage}}
\vspace{1em}

\subsection{Case study Phase}

In this initial stage, we embarked on three separate case studies within established UK based cyber security organisations, denoted as A, B, and C. We conducted semi-structured interviews with cyber security practitioners from these organisations \textbf{(these were teams from organisations)}. Below are the details of organisations A, B, and C, including their domains and demographics.

\begin{itemize}
\item 	Team from organisation A operates in the Incident Response domain and has 7 male employees and 0 female employees. We assigned them the code names A1, A2, A3, A4, A5, A6, and A7.

\item 	Team from organisation B operates in Information Security, with 1 male and 1 female employee. We assigned them the code names B1 and B2.

\item 	Team from organisation C, involved in Incident Response. It has 4 male employees and 1 female employee,  we assigned them the code names C1, C2, C3, C4 and C5.
\end{itemize}

Our aim was to meticulously scrutinise and chart the knowledge profiles of individual teams. Through this process, we iteratively refined and enhanced our knowledge profiling framework. The semi-structured interviews also aimed to elicit participants' coverage of \ac{CyBOK} knowledge as covered under their respective roles, certifications, qualifications and the training acquired. The main themes of interviews were the following:

\begin{itemize}
\item Participants' current role within their organisation.  
\item The qualifications they have.
\item The unique certifications they currently hold, and their significance in supporting their roles within their organisation.
\item Other certifications they hold, but which are not required for their current role.
\item The on-the-job training to support their roles.
\end{itemize}

\begin{figure*}[ht]
  \centering
  \begin{subfigure}[b]{0.4\textwidth}
    \includegraphics[width=\textwidth]{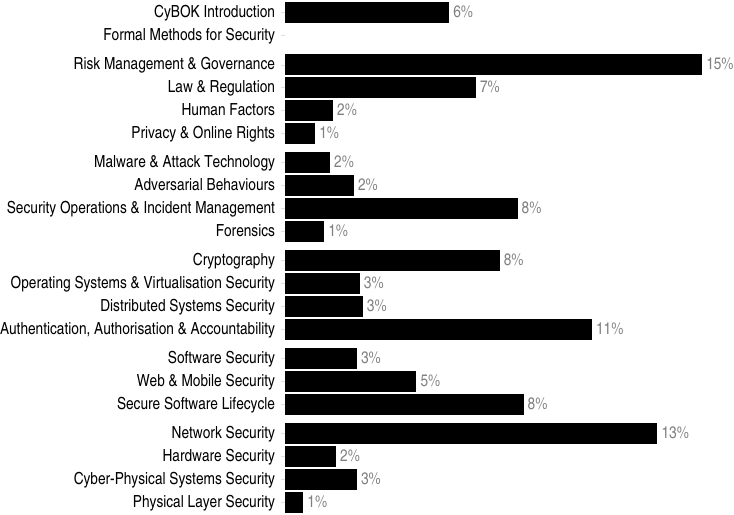}
    \caption{Histogram of Organisation A }
    \label{fig:histoorgA}
    \end{subfigure}
    \hspace{3cm}
\begin{subfigure}[b]{0.4\textwidth}
        \includegraphics[width=.7\textwidth]{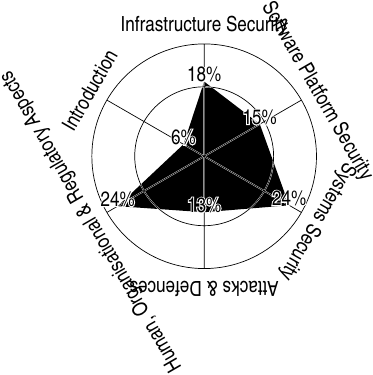}
    \caption{Spider Diagram of Organisation A }
    \label{fig:spiderorgA}
    \end{subfigure}
    \hspace{3cm}
    \caption{Histogram and Spider Diagram of Organisation A to show the \ac{CyBOK} \ac{KA} and \ac{CyBOK} Broad Category Coverage}
\end{figure*}

\begin{figure*}[ht]
  \centering
  \begin{subfigure}[b]{0.4\textwidth}
    \includegraphics[width=\textwidth]{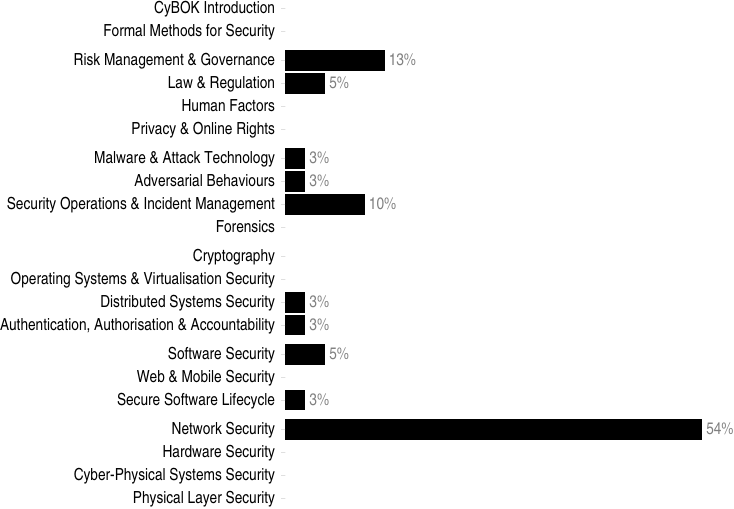}
    \caption{Histogram of Organisation B}
    \label{fig:histoorgB}
    \end{subfigure}
    \hspace{3cm}
\begin{subfigure}[b]{0.4\textwidth}
        \includegraphics[width=.7\textwidth]{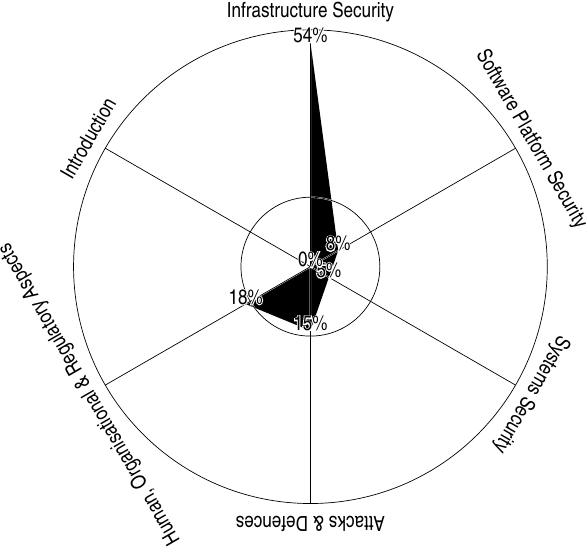 }
    \caption{Spider Diagram of Organisation B }
    \label{fig:spiderorgB}
    \end{subfigure}
    \hspace{3cm}
    \caption{Histogram and Spider Diagram of Organisation B to show the \ac{CyBOK} \ac{KA} and \ac{CyBOK} Broad Category Coverage}
\end{figure*}

\begin{figure*}[ht]
  \centering
  \begin{subfigure}[b]{0.4\textwidth}
    \includegraphics[width=\textwidth]{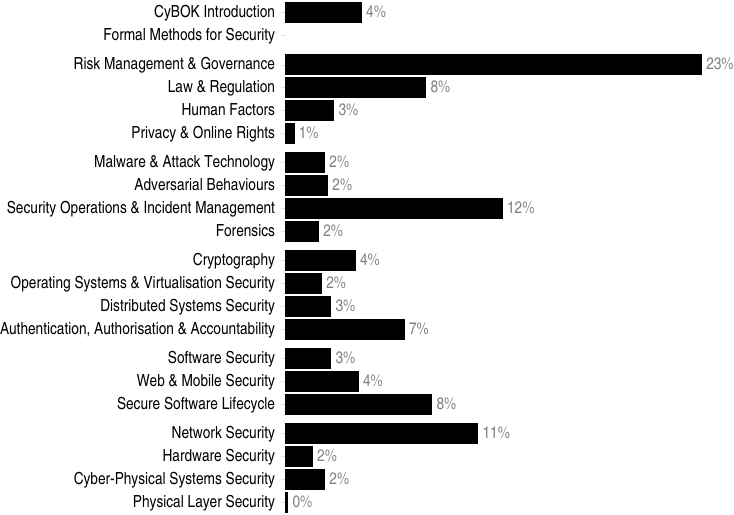}
    \caption{Histogram of Organisation C}
    \label{fig:histoorgC}
    \end{subfigure}
    \hspace{3cm}
\begin{subfigure}[b]{0.4\textwidth}
        \includegraphics[width=.7\textwidth]{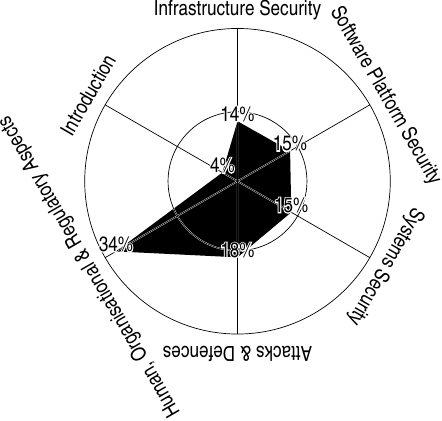}
    \caption{Spider Diagram of Organisation C }
    \label{fig:spiderorgC}
    \end{subfigure}
    \hspace{3cm}
    \caption{Histogram and Spider Diagram of Organisation C to show the \ac{CyBOK} \ac{KA} and \ac{CyBOK} Broad Category Coverage}
\end{figure*}

\subsubsection{Study Participants }

All the participants were recruited through our professional network, and word of mouth. During the course of four (4) weeks, we received fourteen (14) responses from three (3) different organisations. We invited all of them to take part in the study. Participants sample included two (2) females and twelve (12) males. Although all the organisations were dealing with cyber security in the broader aspect, the participants we interviewed worked in specific domains.

\subsubsection{Data Safety and Ethics} 
Before conducting interviews, all participants were informed about the confidentiality of their data and were given the option to end the interview or withdraw consent at any time. Written consent was obtained from each participant prior to the interview. To ensure anonymity, interviews were anonymised before analysis. Participants were provided with a Participant Information Sheet, and Consent forms prior to the interview. These documents are included in the appendix as Figure~\ref{fig:PIS_case},  and ~\ref{fig:Consent_Case}. Approval for this study was granted by the University of Bristol, Faculty of Engineering Research Ethics Committee. Due to the sensitivity of the participants’ work and also to mitigate against any profiles being utilised by their organisation as a performance mechanism, we are unable to share individuals’ data or the dataset in line with requirements from the Ethics Committee.

\subsubsection{Data Collection Method}
All the interviews took place online using Microsoft Teams. We started the interview by first obtaining the signed informed consent from the participants and an agreement for recording to take place. The main interview started with a short introduction of \ac{CyBOK} and the purpose of the knowledge profiling study. Participants were then prompted with a series of questions regarding their experience, roles, training, and certifications. These questions encompassed topics like: 

\texttt{"Could you describe your current role within your organisation and the tasks you undertake?"}

Subsequent questions were tailored based on their responses, delving deeper into related areas. The interviews lasted between 25 and 35 minutes. 

\subsubsection{Data Analysis}
The audio recordings were transcribed using a GDPR compliant transcription service before being analysed. All the transcripts were subsequently rechecked, read, and reread to ensure credibility. The analysis was completed according to respective organisations\footnote{Organisations and participants were not asked for feedback to preserve anonymity of participants in line with our ethics protocol}. Our analysis relied on the detailed questions we asked the participants during the interviews. Then, we carefully sorted their answers into four categories: responsibilities (RES), experience (EXP), certifications (CER) and training (TRA). We followed the steps outlined in Section ~\ref{sec:profile})  and used a \ac{CyBOK} mapping framework ~\cite{nautiyal2022uk} to map these concepts to \ac{CyBOK}. This process was done systematically to make sure we accurately mapped the information, especially when there were no existing mappings available.

\subsubsection{Result - Knowledge profiles of case study organisations} 
\label{sec:case}
We first present some example knowledge profiles of employees (Figures~\ref{fig:HistoX},~\ref{fig:HistoY},~\ref{fig:HistoZ} and~\ref{fig:spiderx}, ~\ref{fig:spidery}, ~\ref{fig:spiderz},) and discuss how the framework enables us to analyse individual employees' knowledge. We do not list employees' organisation reference or their role to ensure participants' anonymity. 
We then  present the cumulative knowledge profile of organisations A, B and C and discuss how the resulting profiles relate to organisational functions.

\textit{Employee X}. This profile falls mainly within the Risk Management \& Governance and Security Operations \& Incident Management \acp{KAs}. The employee had done \ac{CISSP} and \ac{CISM} certifications for which we had pre-existing mappings. They noted that they were an ISO27K lead auditor, so we mapped the concepts from that. They had also done training on advance intrusion detection system concepts, so we mapped that to network security and intrusion detection and prevention systems. They said that they understood quite a bit about domain name system security, so we again mapped that to the relevant concept within network security. They also mentioned that they had knowledge on cyber threat intelligence which we also mapped. Overall, the profile shows a strong understanding of Risk Management \& Governance along with relevant \ac{KA} such as Security Operations \& Incident Management, Network Security and Law \& Regulation. 

\textit{Employee Y}. 
This profile falls mainly within the Secure Software Lifecycle, Network Security, Authentication, Authorisation \& Accountability, Security Operations \& Incident Management and Risk Management \& Governance \acp{KAs}. This shows that Employee Y has a broad understanding of the concepts in these topics. 
Employee Y's profile also reflects expertise in security engineering, mainly secure software development, networking and hardware security  related knowledge. 

\textit{Employee Z}. 
This profile falls mainly within the Authentication, Authorisation \& Accountability, Network Security, Cryptography and Risk Management \& Governance \acp{KAs}. Employee Z's profile indicates a range of applied cryptography expertise, in a range of application contexts such as network security as well as authentication and authorisation.

\textbf{\textit{Cumulative knowledge profile for Organisation A.}} 

When we consider the cumulative knowledge profile of Organisation A (Figures~\ref{fig:histoorgA}, \ref{fig:spiderorgA}), we note strong coverage of knowledge on Security Operations \& Incident Management, Authentication, Authorisation \& Accountability, Software Security, Law \& Regulation, etc. It is clear that the organisation has capacity in this team in security operations, incident response and related topics. However, it is not clear how the knowledge on Forensics and Malware \& Attack Technologies is resourced. It is likely that this is the purview of another team or outsourced. The knowledge profile enables the organisation to reflect on this and whether this coverage is sufficient or if it needs to be addressed through strategic hiring or training. 


\textbf{\textit{Cumulative Knowledge Profile for Organisation B.}} 

When we consider the cumulative knowledge profile of Organisation B (Figures~\ref{fig:histoorgB}, \ref{fig:spiderorgB}), given its domain of Information Security, we would expect to find a strong coverage of the Network Security \ac{KA} focused on securing the organisational network. We also note coverage of relevant knowledge areas on Risk Management \& Governance and Security Operations \& Incident Management. As this was a small team, it is not clear if there is knowledge in other teams on Human Factors (an important consideration for Information Security) as well as Law \& Regulation. It is likely that this knowledge is drawn through other teams within the organisation (e.g., a legal team covering law and regulation). However, the knowledge profile provides a means of reflecting on these aspects and ensuring that they are either covered through other units or improving coverage through investment in staffing or training (or both).

\textbf{\textit{Cumulative knowledge profile for Organisation C.}} 

When we consider the cumulative knowledge profile of Organisation C (Figures~\ref{fig:histoorgC}, \ref{fig:spiderorgC}), given its domain of Incident Response,  we would expect to find a strong coverage of Security Operations \& Incident Management along with Risk Management \& Governance and Network Security. This is reflected strongly in the overall profile. We note the low coverage of Malware \& Attack Technologies, Adversarial behaviour and Forensics. Similar to organisation B, we anticipate that these activities are either the purview of a different unit within the organisation or outsourced. Regardless the knowledge profile provides a means to reflect on these aspects and ensure that they are appropriately covered through other mechanisms (or the need to develop capacity in these \ac{KA}).

\vspace{1em}
\fbox{\begin{minipage}{24em}
\textbf{Cumulative knowledge profile for Organisations:}
As mentioned in Section~\ref{sec:profile}), the cumulative profiles of an organisation's employees collectively shape the organisation's profile. The Knowledge Profile of the Organisation is denoted as $KP_{ORG}$  and is formally characterised as the union of all individual employee knowledge profiles, represented as $KP_{EMP}$, that collectively constitute the organisational knowledge profile. It can be defined as:
\[KP_{ORG} = \bigcup \{KP_{EMP1}, KP_{EMP2}, \ldots KP_{EMPn}\}\]

\end{minipage}}
\vspace{1em}

\subsection{Workshop Phase}

The second phase of this study involved a workshop with cyber security industry professionals. It was designed in order to \emph{test} the initial framework with practitioners and refine it based on their input and feedback. We scheduled two workshops to enable participants have a choice in timing.

\subsubsection{Study Participants }
During the workshop phase, we enlisted participants primarily through our professional network and by word of mouth. Our goal was to gather cyber security experts with experience in diverse security projects. Despite receiving initial responses from 14 individuals, only 8 ultimately participated, with a gender distribution of 3 females and 5 males. Notably, these participants were not associated with the organisations examined in our case studies. They encompassed a range of roles within the industry, including vulnerability analysts, training providers, and senior managers.

\subsubsection{Data Safety and Ethics} 
Before the workshop, participants were assured of the confidentiality of their data and had the choice to leave or withdraw consent anytime. Written consent was obtained from each participant beforehand. They received a Participant Information Sheet, and Consent forms prior to the workshop. These materials can be found in the appendix as Figure~\ref{fig:PIS_work}, and ~\ref{fig:Consent_Workshop}. The study received approval from the University of Bristol, Faculty of Engineering Research Ethics Committee. 

\begin{figure*}[ht]
  \centering
  \begin{subfigure}[b]{0.41\textwidth}
    \includegraphics[width=\textwidth]{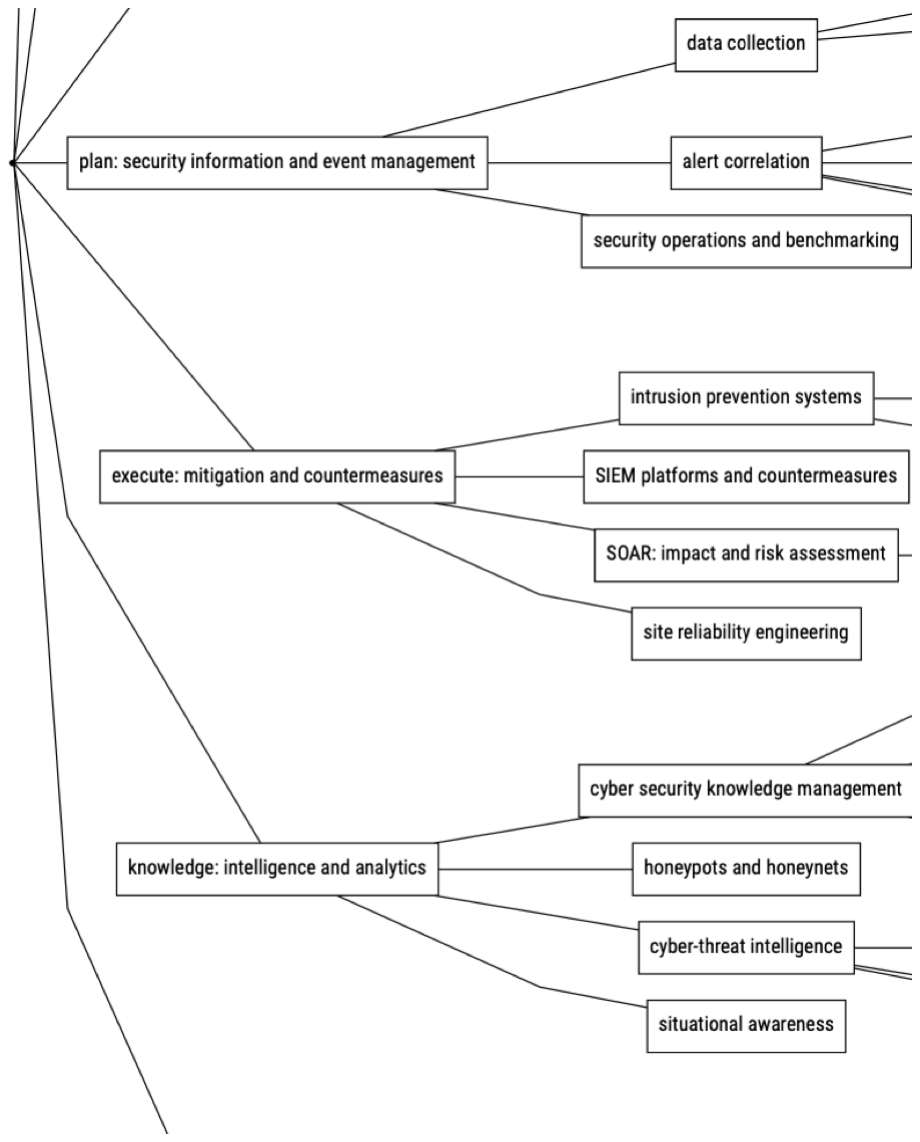 }
    \caption{Total SOIM \ac{KA} Coverage 
 - Organisation  A (Small Part)}
    \label{fig: TreeTotal}
    \end{subfigure}
    \hspace{3cm}
\begin{subfigure}[b]{0.41\textwidth}
    \includegraphics[width=\textwidth] {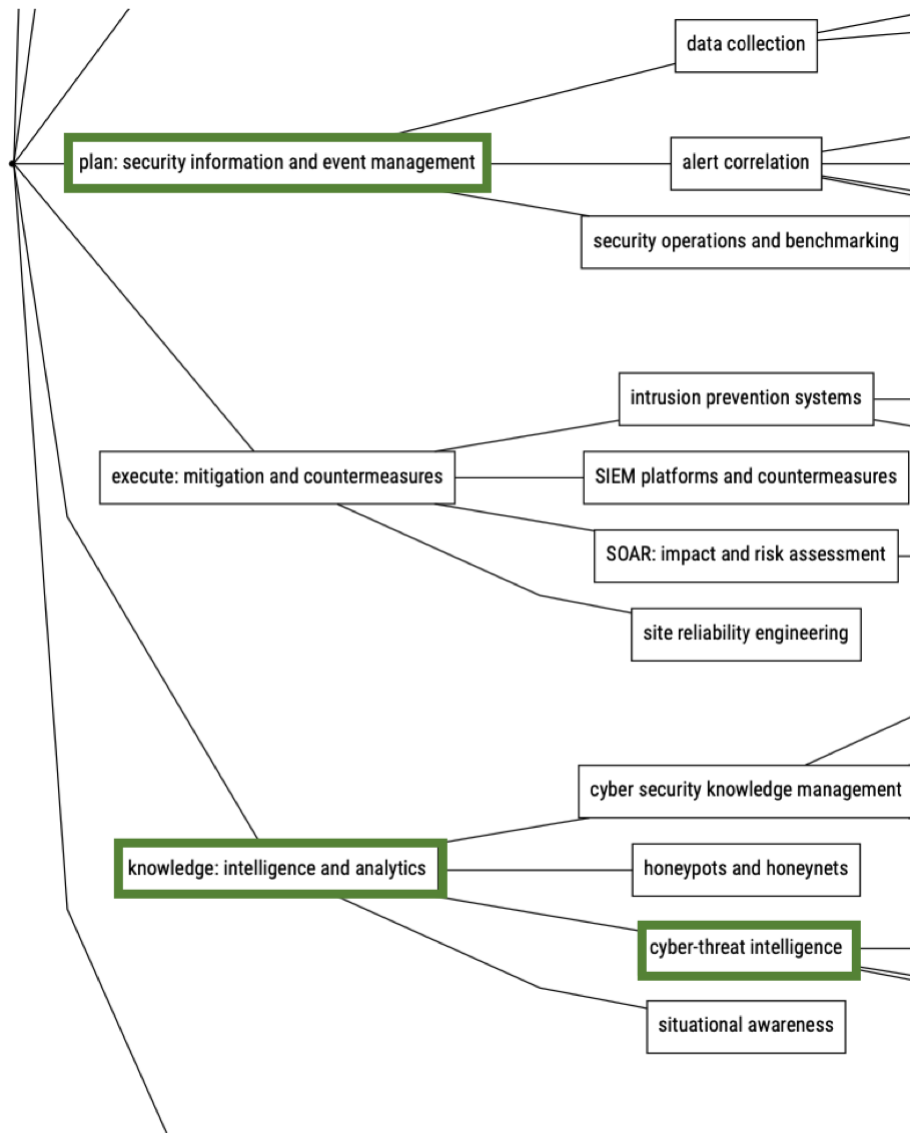 }
    \caption{Practiced SOIM \ac{KA} Coverage - Organisation A (Small Part)}
    \label{fig:TreePracticed}
    \end{subfigure}
    \hspace{3cm}
    \caption{ Visualisation through Tree Diagram (Part of Organisation A) to show the coverage of SOIM \ac{KA} – Total SOIM \ac{KA} coverage and Practiced SOIM \ac{KA} coverage }
\end{figure*}

\subsubsection{Data Collection Method}

We conducted 2 workshops online via Microsoft Teams. Workshop began by presenting the initial knowledge profiles of Organisations A, B, and C. Participants were invited to ask questions and share feedback on the initial \ac{CyBOK} knowledge profiling framework. Presentation of the Knowledge Profiling Framework followed by a group discussion on the following points:

\begin{itemize}
\item What level of granularity would be informative (e.g. spider chart / bar chart / KA topics)?
\item	The cost and effort in deriving an initial knowledge profile.
\item	Considerations around keeping a knowledge profile up to date.
\item	Where/if a deeper granularity is used, is there a CMM-like conceptualisation of levels of knowledge profiles?
\end{itemize}

Participants were also encouraged to raise further points and the discussion took an interactive approach. Participants' permission to video-record the workshop was obtained, so both the sessions of the workshop were video-recorded. 
 
\subsubsection{Thematic Analysis} 
Audio files were transcribed, and notes were consolidated, and we then applied thematic analysis, assigning codes to responses \cite{braun2006using}, \cite{nowell2017thematic}.  The first author identified 14 different codes, covering topics like future certificate updates and using templates. After coding, the first author looked for patterns in the data to understand the main themes. This helped them find five major themes that summarised the discussions. To ensure accuracy, the first author discussed their findings with the second author, who provided additional insights and acted as a reviewer. 

Initially, a total of five themes were identified. 
\begin{itemize}
  \item   Presentation Level 
  \item	Practiced Knowledge 
  \item	Updating of Certification
  \item	Use of Template
 \item	Automated Process for Big Organisations
\end{itemize}

After the second author's review, five themes were merged into the final two themes.
\textbf{Presentation Level} and \textbf{Representation of Currently Practiced Knowledge}.

\textbf{Presentation Level :} 
This theme was composed of three themes: \textit{Presentation Level, Use of Template}, and \textit{Automated Process for Big Organisations}. Stakeholders expressed a desire to view presentations in various formats across these themes, including at the HR level, for third-party or outsource stakeholders, or whether a specific template was employed in the presentation, or if the entire process was generated through an automated tool.

{\textbf{Representation of Currently Practiced Knowledge :}
This theme was composed of \textit{Practiced Knowledge} and \textit{Updating of Certification} themes because stakeholders sought to ascertain whether employees' current knowledge is being updated through new certifications or training, and whether they are effectively applying this knowledge in their current roles.

These themes encapsulate key insights gleaned from the workshop discussions, shedding light on pertinent aspects of cyber security practices within the participating organisations.

\begin{enumerate}

\item {\textbf{Presentation level}}

Participants observed that various presentations would cater to the needs of different decision-makers within an organisation. For example, these could include presentations tailored for the board level, manager level, HR level, or for third-party/outsource stakeholders within the organisation. For instance, one participant noted:

\vspace{1em}

\texttt{"What I come up against very often, and you’ll see this, this is a joke amongst cyber security professionals, that you can’t get the board interested or they think that the IT director has got all of those skills and if you gave something like this to them, they would just tick all the boxes and that’s to a point about inflation of what people say they know."}
\vspace{1em}

There was general feedback that the spider charts would be useful for board-level decision makers while the histograms would be suitable for senior managers. On the other hand, team leaders would want to know details and depth of the knowledge within their team, so annotating and highlighting relevant paths within the Knowledge Tree Figure~\ref{fig: TreeTotal} and ~\ref{fig:TreePracticed} of a \ac{KA} would be most appropriate to show the details with depth of the knowledge, as it is showing  the deeper granularity of details of SOIM \ac{KA}. 

Another  participant noted: 

\vspace{1em}
\texttt{"I think user friendliness of this tool will need to be considered quite carefully. My question is about the audience. Who are you directing this to? Is it the heads of cyber security teams? Is it learning and development? Is it HR? You know, it would be just good to understand who this tool is specifically meant for."}
\vspace{1em}

Participants expressed curiosity regarding the potential benefits of utilising different visualisations at various organisational levels. They were interested in understanding how such visualisations could meet the specific needs of different stakeholders, such as HR professionals or team leaders who might benefit from gaining deeper insights into employee granularity.

Besides, participants were quite keen to know two things: For whom is the Knowledge Profiling being conducted, and what purpose will it serve?

\begin{itemize}
\item {Knowledge Profiling for whom?}

The participants were eager to learn for whom knowledge profiles would be created in the first place, and who would put it to good use. Will it be beneficial for the employee or the management? 
Participants wondered if knowledge profiling could help them show how valuable they are, especially in big organisations where people often stay in the same jobs for a long time. They wanted to know if this profiling could help them move up in their careers. They were worried that if their bosses found out about their weaknesses through these profiles, it could put their jobs at risk. They talked about how knowledge and qualifications should match up for specific jobs. 

They were afraid that not having the right qualifications could hurt their chances of getting or keeping a job. They also asked if having a knowledge profile of their organisation could help their managers see how good they are at their jobs and help them get better. Some were even interested in joining training programmes based on these knowledge profiles.

\item {Knowledge Profiling for what?}

Participants also showed interest in understanding the function that the knowledge profile will play for their organisations. 
Participants wondered if knowledge profiling could make their organisations stand out more in the market by showing off their strengths better. Some thought that if they looked closely at everyone's knowledge, they might find hidden talents that could deliver benefits. Also, the people in charge thought that if they looked at everyone's knowledge closely, they could see where there are gaps in the team and decide better who to hire in the future. They also wondered if this could help them make new rules for hiring people. They talked about whether knowledge profiling could help their organisation use its resources in the best way. They noticed that some people might have knowledge capability that they are not using properly, so by looking at everyone's knowledge closely, they could make sure everyone's talents are being used well, which would help the organisation use its resources better.

\end{itemize}

\item {\textbf{Representation of Currently Practiced Knowledge}} 

Participants noted that often certifications or qualifications can be historic and do not reflect the current or recent roles of an individual. This may lead to an inaccurate representation where an organisation may perceive to have certain knowledge within its workforce but that knowledge may have declined or eroded due to it not being practiced for many years. It was highlighted that the visualisations should represent the level of \emph{currently practiced} knowledge. For instance, one participant noted:

\vspace{1em}
\texttt{"We obviously track people's development and there's hundreds and thousands of people that sit within my organisation, where their career and their training is tracked and dated. They have a competency against each training piece, which is useful for us to be able to map somebody's knowledge and the date of that knowledge as well. So, we can understand how long ago they may have been in a posting, so we understand our military capability within the cyber world fairly well."}

In this context, the concept of "currency" becomes paramount, i.e., validity of a knowledge profile at a current point in time $(T_{c})$.  This is important as certifications expire, training can become dated and experience can deteriorate over time if the knowledge isn't practiced. We note that responsibilities remain current so are not considered further for the following discussion.

We introduce the function $v$ which determines the validity period $(T_{d})$ of a Certification (CER), Training (TRA) or Experience (EXP):

\[T\_{d} \leftarrow v(x)~where~x \subseteq \{CER, TRA, EXP\}\] 

When considering the formal specification in terms of current time $(T_{c})$, we redefine $KP_{EMP}$ in our approach as follows to check that $(T_{c})$ (the discrete point in time) falls within the time interval $(T_{d})$:

\[KP_{EMP} \leftarrow \left[ \bigcup\ {K \over {CYB}} \right]{\_T_{_{c}}} \in {\_T_{_{d}}}\]

\subsection{Threats to validity}

\subsubsection{Threats to Internal Validity}

During the case study phase, we collected data through semi-structured interviews and analysed their responses by mapping it to \ac{CyBOK}. In our \ac{CyBOK} repository, we had already mapped certain certifications like \ac{CISSP}, \ac{CISM}, \ac{SSCP}  etc. For participants with these certifications, we retrieved their results from the repository. However, some individuals mentioned certifications were not part of our repository, such as CISCO, ISO 27001, SAN - GCED, etc. We mapped the table of contents for those certifications. Typically, we conduct this mapping process by using the table of contents. Normally two researchers conduct the mappings and inter-rater reliability is checked. For these additional certifications, only one researcher did the mappings, so we did not test for inter-rater reliability. 

\subsubsection{Threats to External Validity}

Our entire analysis and the proposed \ac{CyBOK} knowledge profiling rely heavily on the information provided by interview participants. We asked about their roles, experiences, training, and certifications, forming the foundation of our analysis. However, if participants from these organisations didn't provide accurate information, it could impact our results and analysis. We had to trust their responses, and inaccuracies could lead to differences in individual or organisational knowledge profiles. Therefore, we acknowledge that there may be potential threats to validity as a result of this.

\vspace{1em}
\fbox{\begin{minipage}{24em}
\textbf{Note:}
The data taken from the interview process has been carefully mapped using the \ac{CyBOK} Mapping Framework~\cite{nautiyal2022uk}, which serves as the basis for mapping degree programmes and professional courses to CyBOK. Currently, there are more than 50 \ac{NCSC} certified degrees across 80 universities in the UK that have been mapped using this process for certification purposes. This is, therefore, a credible means to map expertise, training, certification and responsibilities as part of the knowledge profiling process and makes it possible for other researchers to undertake similar mappings and compare results.

\end{minipage}}
\vspace{1em}

\end{enumerate}

\section{Related Work}\label{sec:related}

According to Fontenele~\etal{}, any cyber security workforce development that ignores the social part of human behaviour on the network is ignoring a vital component of the cyber realm \cite{fontenele2016}. Cultivating talent in the cyber domains, for example, requires understanding that people who are drawn to this domain may have particular social psychological features and inclinations that make them ideally equipped to flourish in this area. Their proposed framework is capable of selecting, ranking, and evaluating the expertise of cyber security professionals. The approach mixes the profile owner's quantitative and qualitative qualities with values obtained from external assessments. 

Our work can potentially complement the social components with the core technical expertise of individuals -- leading to a richer information set for organisations to evaluate. 

Valdez-Almada~\etal{} analyse the unstructured language in 'resumes' to determine knowledge profiles for Software Engineering positions using Natural Language Processing (NLP) and Text Mining (TM) \cite{ valdez2017}. They gleaned data from CVs in order to identify knowledge of software engineers. Their focus is on the person while we expand that with various variables like certificates, training of the individual as well as the organisation as a whole. Our mapping against \ac{CyBOK} helps identify the under-represented knowledge within an organisation which can then be fed to NLP tools to identify suitable candidates to fill those gaps.

Sousa~\etal{}, used a mixed methods approach which included interviews and a questionnaire, modelling employees of two companies\cite{sousa2016}. To develop staff profiles, a factorial analysis was utilised to uncover separate groups, which resulted in specific knowledge profiles. They define many different knowledge profiles - innovators, organisers, and facilitators - and analyse their contribution to innovation using an action research technique. The focus of Sousa~\etal{} was on roles while our focus is on knowledge and using that to build to organisational profiles. Our work can feed into determining roles of individuals pertaining to their knowledge and expertise. 

Knowledge expertise mapping is presented as part of the entire process for the expert conceptual framework by Ismail~\etal{} \cite{ismail2021}. They use publicly available information of individuals like resumes to build their knowledge profile. However certificates, training, responsibilities in current role might not be adequately reflected in public profiles. Our framework can provide the granularity as well as currency in constructing individual knowledge profiles.


Patamakajonpong~\etal{} proposed a framework which is used to look into the specific knowledge and expertise of an expert by using the competency-based approach \cite{patamakajonpong2015}. Knowledge Engineering is implemented for capturing and modelling relevant third parties' experiences. The Capability Maturity Model (CMM) is then used to develop the personnel's capability and maturity level. Bellini explores the impact of CMM certification on organisational learning using the qualitative and quantitative data from a software process improvement effort in which the Italian branch of a global software business was involved from January 1997 to May 2001 \cite{bellini2006}. To explain the rise in productivity in the software development process, a knowledge management viewpoint is adopted. A combination of people and process can contribute to performance. Our work is the first step towards similar rigorous evaluation and capability maturity for cyber security.

By designing and implementing an opinion survey on levels of information taught in universities versus knowledge needed in industry, Garousi~\etal{} examine the knowledge gaps of software engineers\cite{garousi2019}. They created the survey using the SE knowledge areas (KAs) from the most recent version of the Software Engineering Body of Knowledge (SWEBOK v3), which divides SE knowledge into 12 \acp{KAs}, each of which has 67 subareas (sub-\acp{KAs}). Their research is based on (opinion) data from 129 practitioners, the majority of whom are based in Turkey. Our framework is distinct in our focus on cyber security, upon the foundation of \ac{CyBOK} and can potentially lead to a evaluative understanding of individuals/organisations.

An evaluation of professional and occupational profiles based on SWEBOK is undertaken by Quezada-Sarmiento~\etal{}, as well as the development of an ontological model to obtain the necessary information to establish the relationship and the criteria to evaluate professional profiles\cite{quezada2016}. Our work is similar to the work of Quezada-Sarmiento~\etal{} in its study of entities as they are, but also expands to document the evolution of individuals and organisations over time.  

Bourque~\etal{} propose Bloom's taxonomy levels for three software engineer profiles in a software engineering process grouped using SWEBOK topics. The goal of this work is to show how such profiles can be utilised to create job descriptions, software engineering position descriptions inside a software engineering process definition, professional development paths, and training programmes \cite{bourque2003}. Our framework is capable of determining whether a person has sufficient knowledge for their current role, what the organisation's strengths and weaknesses are, and what type of knowledge they will be seeking in a person if they require experts.

Jooss~\etal{} proposed a Model of developing strategic agility through skills-matching.
Their study encompassed 34 interviews conducted with employees from 15 distinct organisations. Their model established a connection between employee talent and the organisation's strategy. It operated on a matching theory that synchronised the skills of employees with the organisation's strategic objectives\cite{jooss2024skills}.
In contrast, our framework prioritised the expertise of employees. Our objective was to construct a comprehensive profile by analysing their historical and current roles, certifications, and training initiatives, rather than exclusively emphasising the organisation's strategies.

\section{Contribution of proposed work in terms of theory and practice}
\label{sec:contri}

Our research on \ac{CyBOK} knowledge profiling has found a big gap in the literature: there isn't much substantial work done on cyber security knowledge profiling. However, we discuss several important ways where our study can help both in theory and practice.

\subsection{In terms of theory}
Current approaches that aim to capture organisational knowledge do so partially. For example, CVs do not always reflect the current state of individuals, nor do they reflect the relevant gaps. For a complete and up-to-date profiling one needs to capture multiple data points about individuals and evaluate them against a living cyber security knowledge base. We present a formal way of capturing diverse experience, role, certificates, and training over time against \ac{CyBOK}. At an aggregate level this reflects the change in organisational knowledge base. Such metrics will help capture the focus of organisations and their expertise, and identify their training needs in their areas of expertise.

\subsection{In terms of practice}
The framework's graphical representation serves as a versatile tool. One can use it to generate various types of graphs. For instance, organisations can easily find the following : 

\begin{itemize}
  
\item An organisation's acquired knowledge through training initiatives.
\item An organisation's knowledge attained through certification programmes.
\item An organisation's knowledge acquired through individuals' current roles within the organisation.
\item An organisation's knowledge gained through experiential learning acquired by individuals within the organisation.
\end{itemize}

Ultimately, it assists organisations in making informed decisions by assessing individuals' knowledge levels, identifying organisational strengths and weaknesses, and pinpointing the expertise required for key roles. In today's business era, outsourcing has become a widely popular aspect of any business. To that end, our framework can help to assess the suitability of contracting organisations. Human resources can make prepare a comprehensive hiring plan for organisations depending on their current knowledge profile and business objectives.


Overall, these findings indicate that knowledge profiling can greatly assist organisations in achieving their goals by ensuring alignment between the right candidate and the right job, as well as selecting the right organisation to fulfil specific tasks effectively. In essence, knowledge profiling serves as a strategic tool for optimising organisational performance and enhancing overall efficiency in task allocation and execution.

\section{Discussion}
\label{sec:Disc}

Having a systematic process to establish and evaluate an organisation's knowledge profile in cyber security opens up a number of key possibilities:

\begin{enumerate}
  
\item \emph{Knowledge profile as a living, evolving view}. An  organisation's knowledge profile is not a static entity. It varies as new employees join, existing ones gain new knowledge and experience or depart. Having a systematic means to trace these changes can enable an organisation to evaluate its positioning in terms of cyber security services, its strengths and areas of improvement. The temporal dimension provides capability to take snapshots of an organisation's knowledge profile at specific intervals (e.g., at points of key strategic shift in products and services) and establish how the organisation's knowledge capability has evolved over time. 

\item \emph{A systematic approach to upskilling}. 
Knowledge underpins skills. Development of systematic knowledge profiles can enable an organisation to establish where employees have the underpinning knowledge to become skilled in additional tools and techniques, hence not only providing opportunities for professional growth to individuals but also extending organisational capability. This also opens up potential to identify areas where knowledge profiles in different organisational units can be brought together to create new capability and offer innovative cyber security products and services .

\item \emph{Addressing the cyber workforce gap.} There is a need for more skilled professionals in the field as highlighted by various surveys, e.g., the annual cyber security workforce studies by ISC2~\cite{workforce_study}. However, this gap cannot be plugged by a singular type of cyber security professional. Different areas within cyber security require different knowledge capability and skills that build on that knowledge capability. A knowledge profile helps organisations identify where to add depth, i.e., more people with similar knowledge but perhaps different skills, and where to add breadth: bringing people with complementary knowledge together. This provides a more strategic approach to developing and/or enhancing organisational capability and addressing workforce gaps. This also opens up the potential for specifying specific knowledge that is of core focus during: recruitment, commissioning of externally provided training or development of bespoke training programmes to address the gaps.

\end{enumerate}

\section{Conclusion}
\label{sec:conc}
In the context of our research, we have introduced a novel knowledge profiling framework tailored specifically for organisations. Our approach utilises \ac{CyBOK} as a standardised reference point for comparative analysis. The development of this framework was a rigorous process that involved multiple phases, including three comprehensive case studies and subsequent refinement through interactive workshops.
The resulting framework serves as a versatile tool for evaluating and understanding an organisation's knowledge landscape from various angles. It offers two distinct analytical perspectives:

\begin{itemize}
    
\item  \emph{Coverage of Knowledge Domains:} 
This dimension focuses on the extent to which an organisation's employees possess expertise in specific knowledge areas. To assess this, we consider factors such as formal education, training, and the practical experiences and responsibilities of employees. By examining these elements, we gain insights into the breadth and depth of knowledge within the organisation, allowing us to identify strengths and potential gaps.

\item \emph{Currency of Knowledge:} 
This dimension emphasises the relevance and up-to-date nature of an organisation's knowledge assets. Depending on the viewpoint, this could relate to how well the knowledge is applied in current roles, projects, or operations. Assessing the currency of knowledge provides valuable insights into an organisation's adaptability and responsiveness to evolving industry trends and technologies.
By applying these two analytical lenses, our framework provides a comprehensive view of an organisation's knowledge profile. This holistic perspective enables decision-makers to make informed choices regarding talent development, resource allocation, and strategic planning. Moreover, it empowers organisations to align their knowledge assets with their strategic goals and ensures that they remain competitive and agile in an ever-changing business landscape.

\end{itemize}

However, the knowledge profile captured can enable a range of additional analytical lenses, e.g., the extent to which an organisation's knowledge capacity building utilises formal certifications and university degree programmes as compared to on-the-job training. Alternatively, one can explore how frequently certifications are kept up-to-date and utilise this as a basis to develop a rolling programme of knowledge profile refresh and updates. As discussed in our motivation, the knowledge profile also enables an organisation to identify if it is meeting its business needs and whether third party suppliers and services have the requisite knowledge to deliver requisite knowledge capability. 

In this study, we acknowledge certain limitations of our work. Primarily, the current \ac{CyBOK} mapping process employed is manual, requiring significant time investment. Furthermore, the limited sample size facilitated a comprehensive examination of individual organisational knowledge profiles. However, scaling up to larger organisations would entail a substantial increase in resource expenditure due to the intensive nature of the interview process and subsequent mapping activities.
In future work, we intend to transition towards automated \ac{CyBOK} mapping methodologies. Departing from traditional interview-based approaches, we propose the utilisation of a structured questionnaire format. This questionnaire will solicit responses from employees pertaining to their experiential background, roles, certifications, and training. Subsequently, leveraging our automated method, we aim to conduct a more streamlined and efficient evaluation of these responses. 

\section*{Acknowledgment}

This work was supported by the United Kingdom's National Cyber Security Programme.

\bibliography{bibliography}

\clearpage
\appendix
\vspace{10mm}
\noindent\begin{minipage}{\textwidth}
\centering
\includegraphics[height=0.9\vsize, width=\hsize]{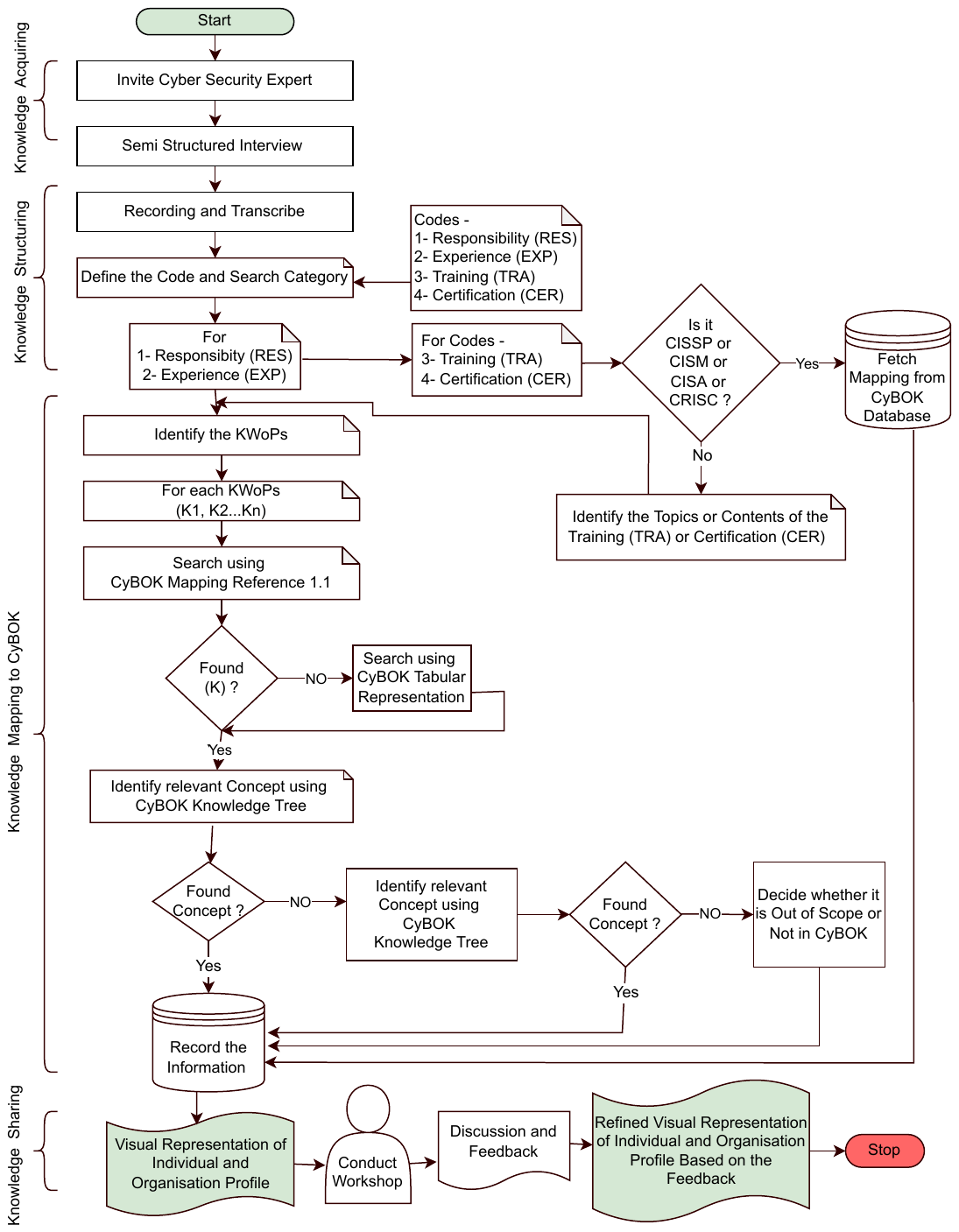} 
\vspace*{-4cm}
\captionof{figure}{Detailed Methodology Flow Chart}
\label{fig:FlowChart}
\end{minipage}

\clearpage
\appendix

\noindent\begin{minipage}{\textwidth}
\centering
\includegraphics[height=0.95\vsize, width=\hsize]{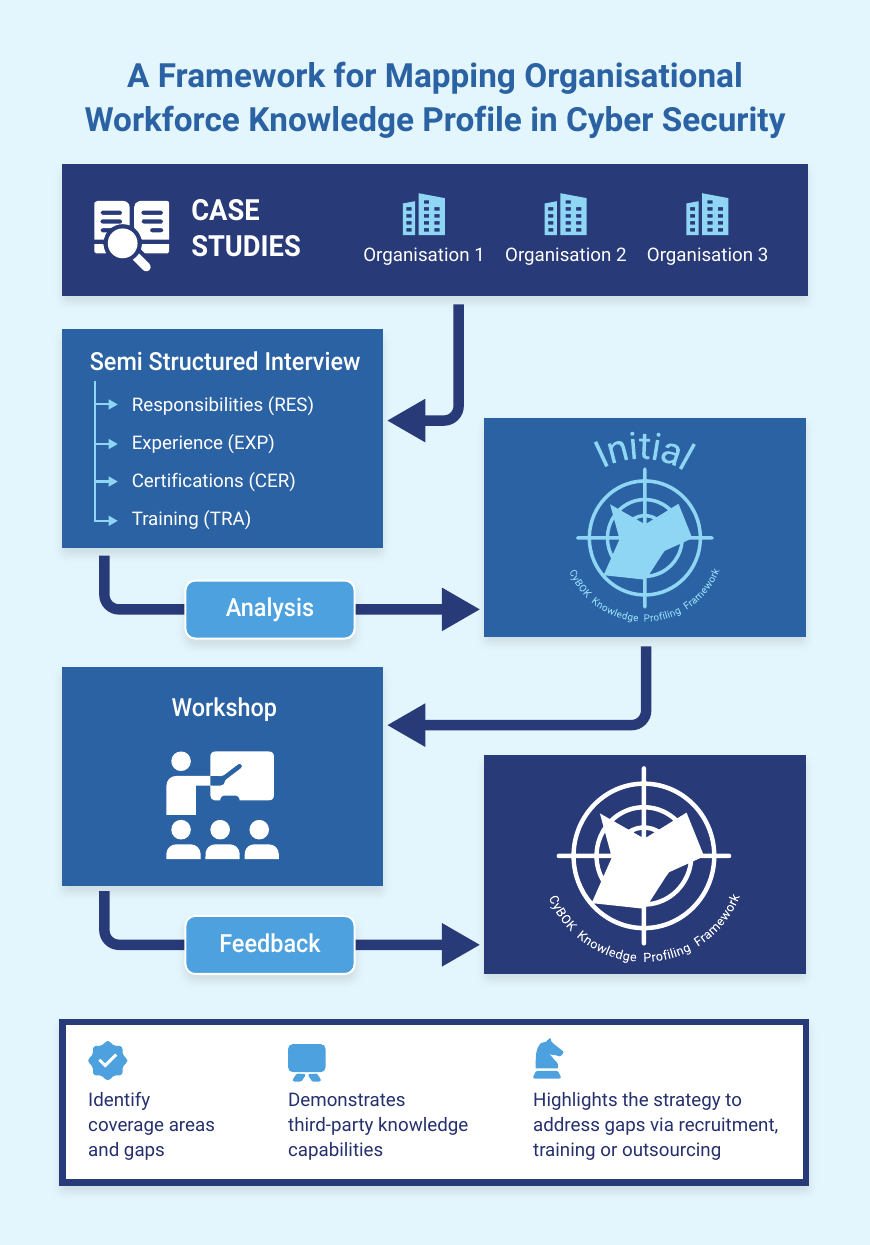} 

\captionof{figure}{Graphical Abstract}
\label{fig:Graphical_Abstract}

\end{minipage}

\clearpage
\appendix

\noindent\begin{minipage}{\textwidth}
\centering
\includegraphics[height=0.95\vsize, width=\hsize]{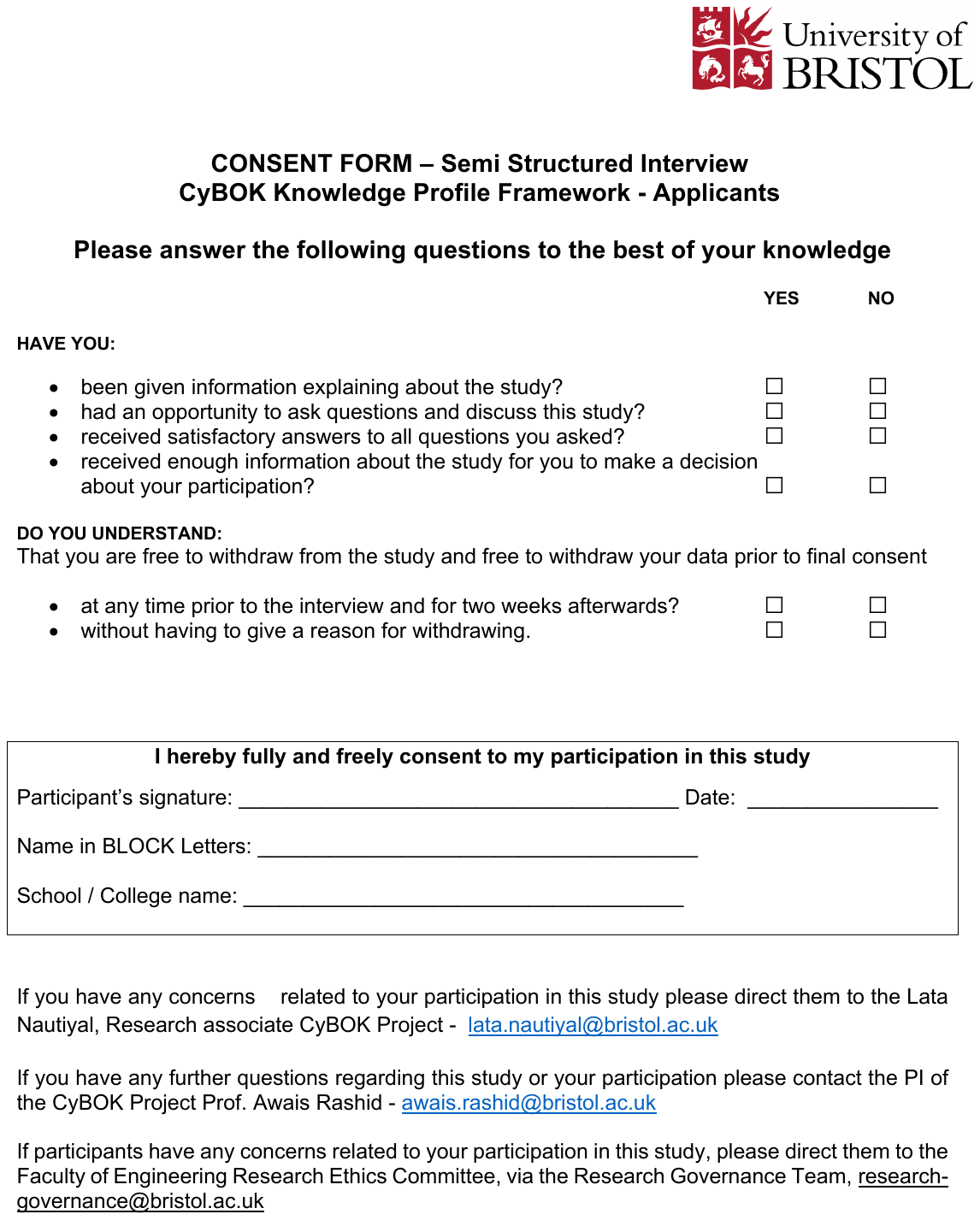} 

\captionof{figure}{Consent Form - Case Study}
\label{fig:Consent_Case}
\end{minipage}

\clearpage
\appendix
\noindent\begin{minipage}{\textwidth}
\centering
\includegraphics[height=0.95\vsize, width=\hsize]{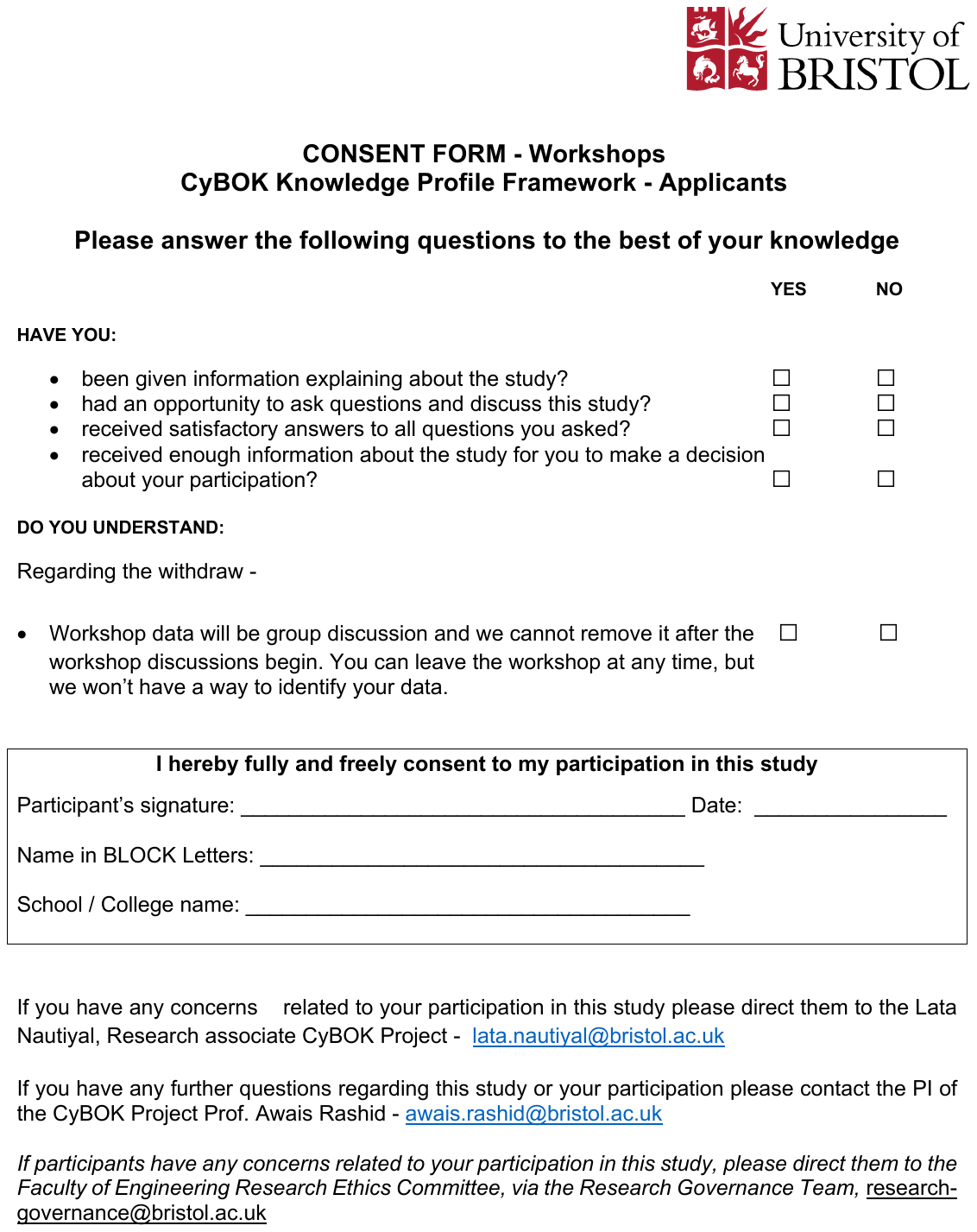} 
\captionof{figure}{Consent Form - Workshop}
\label{fig:Consent_Workshop}

\end{minipage}

\clearpage
\appendix
\noindent\begin{minipage}{\textwidth}
\centering
\includegraphics[height=0.9\vsize, width=\hsize]{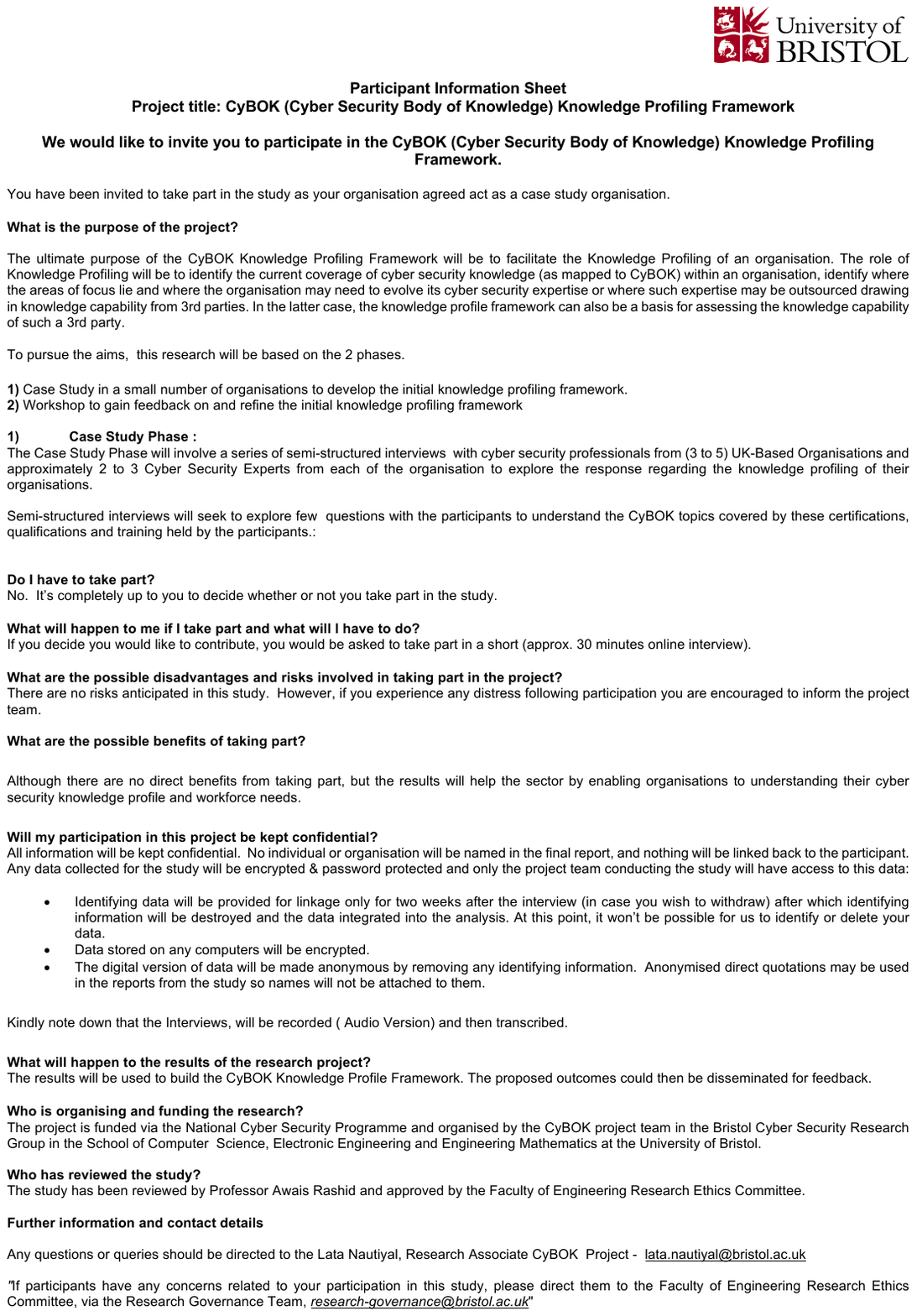} 
\captionof{figure}{Participant Information Sheet -Case study}
\label{fig:PIS_case}
\end{minipage}

\clearpage
\appendix
\noindent\begin{minipage}{\textwidth}
\centering
\includegraphics[height=0.9\vsize, width=\hsize]{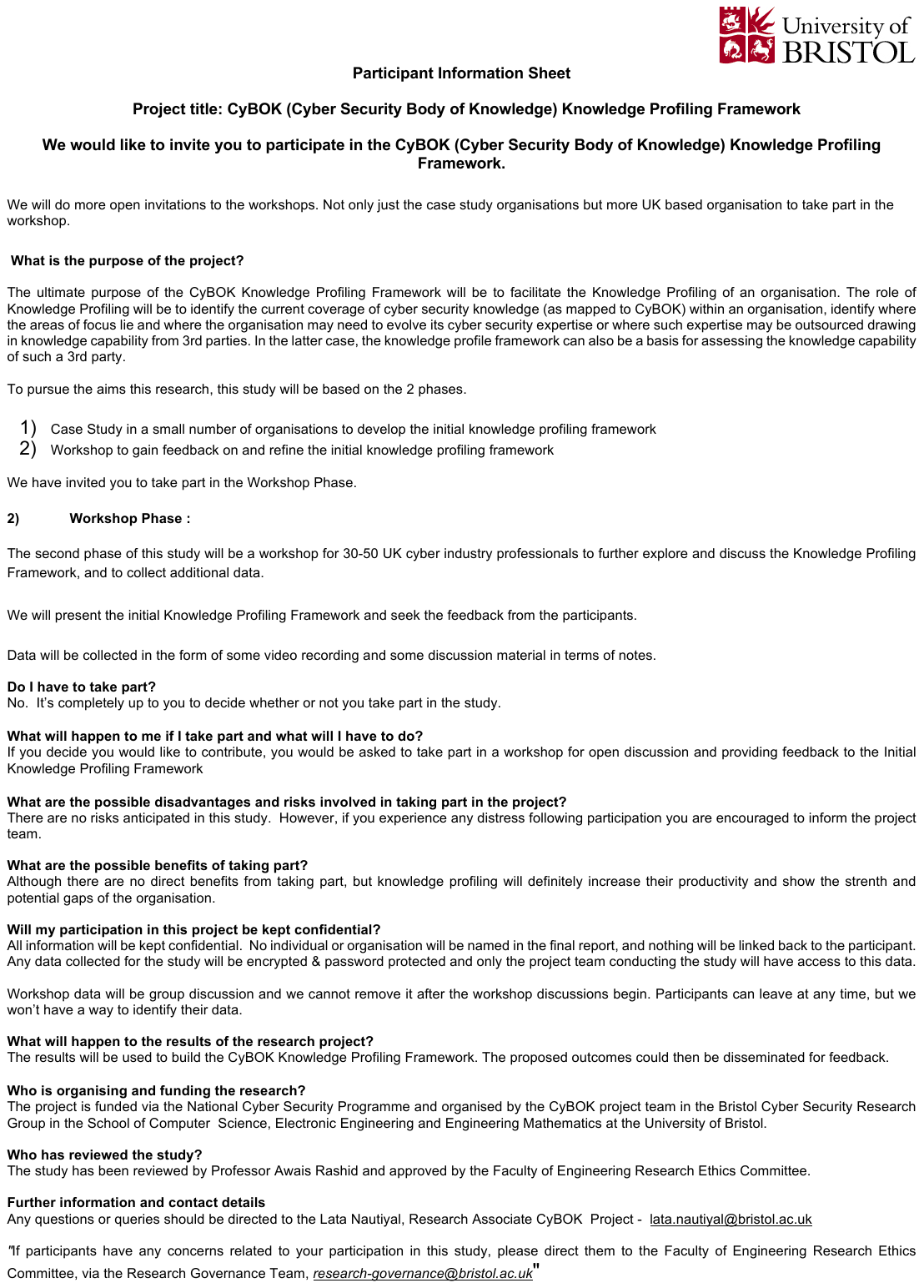} 
\captionof{figure}{Participant Information Sheet - Workshop}
\label{fig:PIS_work}
\end{minipage}
\clearpage
\end{document}